\documentclass[article]{aa} % for a referee version
\usepackage{natbib}         % pour A&amp;A      
\usepackage{graphicx}
\usepackage{epstopdf} 
\DeclareGraphicsExtensions{.pdf,.jpeg,.png,.eps}
\usepackage[switch]{lineno} % switch=Put the line into the inner margin
\bibpunct{(}{)}{;}{a}{}{,} % to follow the A&amp;A style
\usepackage{amsmath}
\usepackage{subfig}
\usepackage{xcolor}		%\textcolor{red}{Text to be colored}
\usepackage{tikz-cd}
\usepackage{url}

\newcommand{\ind}[1]{_{\mathrm{#1}}}
\newcommand{\lsim}{\raisebox{-0.13cm}{~\shortstack{$<$ \\[-0.07cm] $\sim$}}~}
\newcommand{\logg}{\log (g)}

\def\Kepler{\emph{Kepler}}

\def\numax{\nu\ind{max}}
\def\Dnu{\Delta\nu}
\def\deltapi{\Delta\Pi_{1}}
\def\dP{\Delta P}
\def\Eps{\varepsilon}
\def\Teff{T\ind{eff}}
\def\Tref{T\ind{ref}}
\def\Tint{T\ind{int}}
\def\DeltaTeff{\Delta T\ind{ref}}
\def\Msol{M_{\odot}}
\def\ginit{g\ind{init}}
\def\gcal{g\ind{cal}}

\def\MunA{M\mathrm{1}A}
\def\MunB{M\mathrm{1}B}
\def\Mdeux{M\mathrm{2}}
\def\Mtrois{M\mathrm{3}}
\def\Mquatre{M\mathrm{4}}
\def\Mcinq{M\mathrm{5}}
\def\Msix{M\mathrm{6}}

\begin{document}
\title{Red giants evolutionary status determination: the complete $\Kepler$ catalog.
}
%Red giants evolutionary status determination: a comparison between spectroscopic and seismic results
\titlerunning{Red giants evolutionary status determination}
\authorrunning{M. Vrard et al.}
\author{
M. Vrard\inst{1,2}, M. H. Pinsonneault\inst{1}, Y. Elsworth\inst{3}, M. Hon\inst{4}, T. Kallinger\inst{5}, J. Kuszlewicz\inst{6}, B. Mosser\inst{7}, R.~A.~Garc\'\i a\inst{8}, J. Tayar\inst{9,10}, R. Bennett\inst{1}, K. Cao\inst{11}, S. Hekker\inst{12,13}, L. Loyer\inst{1}, S. Mathur\inst{14,15} \& D. Stello\inst{16,17}
% rang 1
} \offprints{mathieu.vrard@oca.eu}

\institute{Department of Astronomy, The Ohio State University, 140 West 18th Avenue, Columbus OH 43210, USA; \email{mathieu.vrard@oca.eu}
\and
Observatoire de la C\^ote d’Azur, CNRS, Laboratoire Lagrange, Bd de l’Observatoire, CS 34229, 06304 Nice C\'edex 4, France
\and
School of Physics and Astronomy, University of Birmingham, Birmingham B15 2TT, UK
\and
Kavli Institute for Astrophysics and Space Research, Massachusetts Institute of Technology, Cambridge, MA 02139, USA
\and
Institute for Astrophysics (IfA), University of Vienna, T\"urkenschanzstrasse 17, 1180 Vienna, Austria
\and
Max-Planck-Institut f\"ur Sonnensystemforschung, Justus-von-Liebig-Weg 3, 37077 G\"ottingen, Germany
\and
LESIA, CNRS, PSL Research University, Universit\'e Pierre et Marie Curie, Universit\'e Denis Diderot, Observatoire de Paris, 92195 Meudon cedex, France
\and
Universit\'e Paris-Saclay, Universit\'e Paris Cit\'e, CEA, CNRS, AIM,  91191, Gif-sur-Yvette, France
\and
Department of Astronomy, University of Florida, Bryant Space Science Center, Stadium Road, Gainesville, FL 32611, USA
\and
Institute for Astronomy, University of Hawai‘i at M$\overline{a}$noa, 2680 Woodlawn Drive, Honolulu, HI 96822, USA
\and
Department of Physics, The Ohio State University, 191 West Woodruff Ave, Columbus, OH 43210, USA
\and
Heidelberg Institute for Theoretical Studies, Schloss-Wolfsbrunnenweg 35, 69118 Heidelberg, Germany
\and
Zentrum f\"ur Astronomie, Landessternwarte (ZAH/LSW), Heidelberg University, K\"onigstuhl 12, 69117 Heidelberg, Germany
\and
Instituto de Astrof\'isica de Canarias (IAC), E-38205 La Laguna, Tenerife, Spain
\and 
Universidad de La Laguna (ULL), Departamento de Astrof\'isica, E-38206 La Laguna, Tenerife, Spain
\and
School of Physics, University of New South Wales, NSW 2052, Australia
\and
Sydney Institute for Astronomy (SIfA), School of Physics, University of Sydney, NSW 2006, Australia
}
%\date{Written on}

\abstract{Evolved cool stars have three distinct evolutionary status: shell Hydrogen-burning (RGB), core Helium and shell Hydrogen burning (RC), and double shell burning (AGB). Asteroseismology can distinguish between the RC and the other status, but distinguishing RGB and AGB has been difficult seismically and spectroscopically. The precise boundaries of different status in the HR diagram have also been difficult to establish.}
{In this article, we present a comprehensive catalog of asteroseismic evolutionary status, RGB and RC, for evolved red giants in the $\Kepler$ field. We carefully examine boundary cases to define the lower edge of the RC phase in radius and surface gravity. We also test different published asteroseisemic methods claiming to distinguish AGB and RGB stars against a sample where AGB candidates were selected using a spectrocopic identification method.}
{We used six different seismic techniques to distinguish RC and RGB stars, and tested two proposed methods for distinguishing AGB and RGB stars. These status were compared with those inferred from spectroscopy.}
{We present consensus evolutionary status for $18 784$ stars out of the $30,337$ red giants present in the $\Kepler$ data, including $11 516$ stars with APOGEE spectra available. The agreement between seismic and spectroscopic classification is excellent for distinguishing RC stars, agreeing at the $94 \%$ level. Most disagreements can be traced to uncertainties in spectroscopic parameters, but some are caused by blends with background stars. We find a sharp lower boundary in surface gravity at $\logg$ $=$ $2.99\pm0.01$ for the RC and discuss the implications. We demonstrate that asteroseismic tools for distinguishing AGB and RGB stars are consistent with spectroscopic evolutionary status at near the RC (with the asteroseismic large separation $\Dnu\leq2\mu$Hz) but that the agreement between the different methods decreases rapidly as the star evolves during the AGB phase.}
{This work present the most complete evolutionary status catalog for $\Kepler$ and APOGEE red giant stars. The data precisely defines the locus of RC stars in the HR diagram, an important constraint for stellar theory and stellar populations. We also demonstrate that asteroseismic tools can distinguish between AGB and RGB stars under some circumstances, which is important for the age estimation for field stars. However, we also put forward the importance of using several techniques to assess the evolutionary status determination for luminous red giants.
}


\maketitle

\voffset = 1.2cm
%________________________________________________________________

\section{Introduction\label{Introduction}}

Luminous red giants are prime targets for modern surveys of Galactic populations and their implications for galaxy formation and evolution. Historically, however, it has proven challenging to infer masses and ages for red giants. The underlying cause is straightforward; stars with a wide range of birth mass and composition converge into a remarkably small range of surface temperature after leaving the main sequence, particularly when they burn their helium in their core. Furthermore, all evolved giants have a wide dynamic range in luminosity. Therefore, both of the classical mass diagnostics on the main sequence (surface temperature and luminosity) are difficult to use as reliable mass or age indicators for cool giants. An important solution to that problem came with the emergence of large scale asteroseismology. With the launch of the space missions CoRoT \citep[Convection, Rotation and planetary Transits:][]{2006ESASP1306...33B} and $\Kepler$ \citep{2010Sci...327..977B}, we discovered that virtually all cool evolved stars exhibit rich power spectra showing evidence for many oscillation frequencies. Turbulence in the outer layers of giants, which excites waves, is a natural physical explanation. In fact, red giant stars are solar-like oscillators that present mostly pressure modes in their spectra. They are the signature of acoustic waves stochastically excited by turbulent convection in the outer layers of the star. In combination with effective temperatures, the information derived from the radial modes is used to deliver unique information on stellar masses and radii \citep[e.g.][]{2010A&A...509A..77K}.
\newline

Red giant stars can also be divided into three distinct evolutionary status. First ascent red giants, hereafter RGB stars, have a hydrogen-burning shell surrounding an inert helium core. Once helium ignites in the core, they become red clump (hereafter RC) stars, with core He burning and shell-hydrogen burning. Once RC stars exhaust core helium, they become double shell burning stars (hydrogen and helium), and they asymptotically approach the RGB; we refer to these as asymptotic red giants, hereafter AGB. The RC is prominent in field stellar populations, and it is a potent diagnostic of stellar populations and stellar physics. As an example, RC stars are commonly used as distance and extinction indicators \citep[see e.g.][ for a more complete description of the use of RC stars]{2016ARA&A..54...95G}.

However, the spectroscopic parameters of RC stars are very close to RGB and AGB stars, therefore producing a possible confusion between the evolutionary status. If the observed sample of stars spans a large range of metallicity, distances or ages, the location of the RC overlaps the RGB in the colour-magnitude diagram, therefore rendering its identification difficult. In addition, RC and RGB stars have very different internal structures, and RC stars can have experienced significant mass loss relative to RGB stars because they are in a later evolutionary status. Both of these effects can be important for understanding the ages and masses of both stellar populations.
\newline

Fortunately, with the advent of asteroseismology it became apparent that RC and RGB stars had different oscillation frequency patterns \citep{2009A&A...506...57D,2012A&A...540A.143M}, tied to differences in their internal structure. These pattern differences, as discussed below, can be used to infer asteroseismic evolutionary status that are precise and accurate. \citep[e.g.][]{2011Natur.471..608B,2011A&A...532A..86M}.  The underlying cause is that red giant oscillation spectra exhibit non-radial mixed modes \citep{2011Sci...332..205B} in addition to pure pressure modes. Because they behave as acoustic waves in the envelope and as gravity waves in the core, they carry unique information on the physical conditions inside the stellar cores. Therefore, dipole mixed modes can be used to distinguish RC stars from RGB stars. Several automated methods were developed during the last decade using differences in the mixed mode pattern to separate RC from RGB stars \citep{2013ApJ...765L..41S,2016A&A...588A..87V,2017MNRAS.466.3344E,2017MNRAS.469.4578H,2018MNRAS.476.3233H}. Several other techniques, focusing on the use of the pressure modes, were also developed \citep{2012A&A...541A..51K,2019A&A...622A..76M}. These methods are supported by evidence that the evolutionary status also has an influence on the pressure mode pattern through the measurement of the Helium second ionization zone \citep{2015A&A...579A..84V,2021A&A...650A.115D} or the solar-like oscillation envelope \citep{2012A&A...537A..30M,2019A&A...622A..76M}. \citet{2012A&A...541A..51K} and \citet{2019A&A...622A..76M} also attempted to distinguish AGB from high-luminosity RGB stars, but without independent verification.
\newline

The precise determination of the evolutionary status also recently allowed characterization of the beginning of the RC phase called the zero-age sequence of Helium burning stars (ZAHB). This feature, a robust prediction of stellar structure and evolution theory, was first clearly identified in globular star clusters and quantified in the 1950s \citep[e.g.][]{1953AJ.....58....4A}. However, it was more difficult to observe in field populations until the advent of asteroseismology. With asteroseismic determinations of evolutionary status, a clear boundary on the mass-radius diagram is evident \citep{2021MNRAS.501.3162L}. Stars below the bulk population were identified as peculiar stars showing evidence of mass loss \citep{2022NatAs...6..673L}. The precise location in the HR diagram of the ZAHB is a sensitive test of stellar model physics. A good example is core overshooting
\citep{2018ApJ...868..150T}. Therefore, measuring the evolutionary status of a higher number of stars will lead to a better understanding of the RC characteristics and the physics occurring during this phase of stellar evolution.
\newline

Spectroscopy and precise photometry can also be used to separate out stars in different evolutionary status. This is most clearly seen in star clusters, where the RC and RGB are distinctively separated, with the RC being systematically hotter. Core He-burning stars with a wide range of masses and metallicities are found in the RC. The temperature offset is a weak function of mass and metallicity. At sufficiently low mass or metallicity a blue horizontal branch emerges, and the temperature offset becomes a strong function of mass; however, such stars are typically too hot to show solar-like oscillations. The AGB is more sparsely populated, but is also hotter, and clearly distinct from the RGB in populous globular clusters. For those stars, it is therefore possible to clearly separate RGB from RC and AGB stars by defining a temperature difference between these populations with composition, mass and age information. These differences are blurred in field populations, where the RGB locus is a very sensitive function of metallicity, and masses are more challenging to infer using traditional tools.

The APOGEE survey \citep{2017AJ....154...94M,2019PASP..131e5001W} used asteroseismic surface gravities as a fundamental calibrator. However, it quickly became apparent that there was an offset between asteroseismic and spectroscopic surface gravities that depends on evolutionary status. This finding implied that it was necessary to infer a spectroscopic evolutionary status for the vast majority of APOGEE targets without asteroseismology in order to ensure that the appropriate correction was applied to the data. The basic principle was as follows. At any given age and composition, there is a clear slope in $\Teff$ as a function of $\logg$ for the RGB in the HR diagram that can be empirically measured. The APOGEE survey provides precise and accurate abundances \citep{2020AJ....160..120J}. For the overlap sample between APOGEE and $\Kepler$, the mass, metallicity, and evolutionary status are all known. The observed properties of the stars can then be used to remove mass and composition effects, producing a temperature offset distribution for field stars that is similar to the star cluster case. When this is done, the temperature offset between the RC and RGB is clearly seen, and can be used as a reliable diagnostic of evolutionary status \citep{2020AJ....160..120J}. This approach was used in APOGEE DR16 \citep{2020AJ....160..120J}, and a related formulation was derived for DR17 \citep{2024AJ....167..208W}.

The process just described did not include any attempt to distinguish between RGB and AGB status. However, the same principles can be applied to them as for the RC: there are two clearly separated branches just above the RC in the HR diagram. The main, crucial difference is that the AGB and RGB converge at higher luminosity and lower surface gravity compared to the RC, meaning that they are not easy to distinguish in the luminous domain.  We discuss our detailed method for separating these groups in Section \ref{AGB_classification}.

\citet{2019MNRAS.489.4641E} used the large number of  automated techniques, both asteroseismic and spectroscopic, able to evaluate the evolutionary status of a large number of red giant stars. Good agreement was found between the methods. We believe that it is appropriate to re-address this question for several reasons. First of all, new techniques have been developed since then \citep{2019A&A...622A..76M,2020MNRAS.497.4843K}. We also have an extended sample available for analysis called the $\Kepler$ Red giant Legacy catalog. This final $\Kepler$ red giant data sample, treated specifically for seismology, were recently analyzed (Garcia et al. in preparation). The $\Kepler$ seismic sample now contains $30,337$ red giants and the APOGEE data release $15,464$, while \citet{2019MNRAS.489.4641E} concentrated their analysis on $6,661$ stars. A larger spectroscopic data set from APOGEE is also available, allowing a thorough study. In this paper, using the extensive amount of data available, we aim at comparing and assessing the reliability of seismic methods to obtain the evolutionary status of red giants (RC or RGB). We will then evaluate the agreement with the spectroscopic evolutionary status determination, thereby improving its calibration. Additionally, we will analyze what these new evolutionary status results bring to our understanding of stellar evolution and stellar models using seismic and spectroscopic data. For this analysis, we will focus on a subsample of stars that have APOGEE data: APOKASC-3 \citep{2024arXiv241000102P}. With our large sample, we can also perform a search of stars below the traditional location of the RC, to precisely measure the minimum surface gravity for the population. Finally, we will discuss the possibility to separate, seismically and spectroscopically, the AGB from the RGB stars beyond the RC.

The layout of the paper is as follows: Section $2$ describes the seismic and spectroscopic data and the extraction of the main seismic and spectroscopic stellar parameters, Section $3$ introduces the characteristics of the different seismic methods that were used in this work, Section $4$ sets out the merging of the methods to obtain the finalized list of stellar evolutionary status and evaluate the agreement between the different methods, Section $5$ details how the spectroscopic classification was performed, Section $6$ is devoted to the comparison between the results of spectroscopic and seismic classification, Section $7$ approaches how the development of such a large sample of stars with identified evolutionary status can be used as observational constraints for models, Section $8$ addresses the determination of the evolutionary status beyond the RC phase, Section $9$ is devoted to our conclusions.

\section{Seismic data selection and description\label{Data selection}}

\subsection{$Kepler$ observations and data selection\label{Data selection}}

%\modif{(For this section, ask Rafael Garcia to describe the data selection, data treatment and global seismic parameters determination)}

We used the KEPSEISMIC\footnote{\url{https://archive.stsci.edu/prepds/kepseismic/}} long cadence \emph{Kepler} light curves. The data were corrected following the methods explained in \citet{2011MNRAS.414L...6G} with gaps filled using a multiscale discrete cosine transform \citep{2014A&A...568A..10G}. The initial seismic sample of stars was obtained from the preparatory work done for the red giant \emph{Kepler} legacy catalog (Garc\'\i a et al. in preparation) containing $30,337$ candidate stars. For each of them, if the presence of oscillations were asserted by Garc\'\i a et al. (in preparation), the global seismic parameters (the large separation $\Dnu$ and frequency $\numax$ of maximum oscillations) were determined following different methods. When at least three of those methods identified oscillations and were able to measure $\Dnu$ and $\numax$, the median values of those two parameters from the different methods were taken as preliminary results. After filtering for outliers and background stars, preliminary $\Dnu$ and $\numax$ were determined for $25,393$ stars (Garc\'\i a et al. in preparation, see also table \ref{Table_example_evolutionary_status_results}). Although we use the ensemble of these stars for the evolutionary status analysis, we later focus on a subsample of $15,464$ stars with APOGEE values from \citet{2024arXiv241000102P}. For the rest of the paper we call this subsample the APOKASC-3 sample..
%In this work, we used preliminary $\Dnu$ and $\numax$ values for only $25,393$ stars (Garc\'\i a et al. in preparation) that correspond to the median values given by the different methods for each target after filtering for outliers and background stars.

\subsection{Determination of the global seismic parameters \label{Global parameters determination}}

\citet{2024arXiv241000102P} inferred the global asteroseismic properties $\Dnu$ and $\numax$ for all targets. They used a total of $10$ distinct techniques for $\numax$ and $7$ for $\Dnu$ and performed a rejection of outliers and background sources. Where possible, background sources were identified with the use of the spectroscopic APOGEE data: an estimation of $\numax$ is deduced from spectroscopy, with surface gravities and effective temperatures, and compared to the measured $\numax$ \citep{2024arXiv241000102P}. If this measurement is out of the rough domain expected given the spectroscopic data, the value is considered as a background source. An outlier rejection processed was also performed by rejecting measurements discrepant from the median at more than $5\sigma$. Details of the process are described in \citet{2024arXiv241000102P} and Garcia et al (in preparation). For stars without APOGEE spectra we required detections in both $\Dnu$ and $\numax$ from $3$ different methods. We then used the median values as an estimation of $\Dnu$ and $\numax$ in the rest of the paper. These estimations were then used as an aid by the different evolutionary status determination methods. We note that the global seismic parameters for the stars without APOGEE spectroscopic data are preliminary ones. We present the evolutionary status of these stars in this paper, and defer a more detailed discussion of their properties to Garcia et al. (in prep).

\subsection{APOGEE data description and selection}

Our paper uses spectroscopic data obtained with the 2.5-meter Sloan Foundation telescope \citep{2006AJ....131.2332G} as a part of the high-resolution APOGEE near-infrared survey \citep{2017AJ....154...94M,2019PASP..131e5001W}. Spectroscopic parameters were inferred with the ASPCAP pipeline \citep{2016AJ....151..144G}, applying the linelists \citep{2015ApJS..221...24S,2021AJ....161..254S} and the calibration procedure described in \citet{2020AJ....160..120J}. 

We use the sixteenth data release from the Sloan Digital Sky Survey \citep{2020ApJS..249....3A}, hereafter DR16. All $\Kepler$ field targets were obtained prior to the most recent data release, DR17 \citep{2022ApJS..259...35A}, so the DR16 sample is complete. In DR16  a polynomial fit to distinguish spectroscopic evolutionary status was adopted, then the asteroseismic status was applied as a calibration sample (see Section \ref{Section:Spectro_Determination_technique} below). This was replaced with a neural net classification in DR17, which is less useful for our purposes, because we need to extend the APOGEE calibration to AGB stars. We therefore work with the DR16 data for this project; the two releases are calibrated to the same fundamental system and are close to one another in mean properties. See \citet{2024arXiv241000102P} for a more detailed comparison.

Our $\Kepler$ field data were selected to prioritize targets with time domain data. See \citet{2014ApJS..215...19P} for a discussion of the $\Kepler$ selection function, \citet{2017AJ....154..198Z} for APOGEE targeting, and \citet{2018ApJS..239...32P} for specific targeting data in the $\Kepler$ fields.

\section{Seismic methods description}

In this section, we will briefly describe the different methods that were used to determine the evolutionary status of red giant stars. Most of these methods focus on the separation between RC (He-core burning objects, including massive stars usually called secondary red-clump stars) and RGB stars because differentiating AGB and high-luminosity RGB stars is particularly challenging. AGB and RGB stars have a similar mode pattern with small $\Dnu$ and $\numax$ and similar gravity-mode period spacings ($\deltapi$) \citep[][their Figure $4$b]{2013ApJ...765L..41S}. Moreover, the increasing lifetime of the mixed modes when the star evolves on the RGB or AGB branch render the detection of those modes particularly difficult with $\Kepler$ length of observation; the decreasing coupling to the core with increasing luminosity makes detecting mixed modes even more difficult \citep{2014A&A...572A..11G}. 
Therefore, in this Section we focused on the separation between RC and RGB stars.

%Put here the description of the spectroscopic data, for Marc to write it.

\subsection{First Method (M$1$): \citet{2016A&A...588A..87V}}

This first method is based on the measurement of the asymptotic gravity-mode period spacing ($\deltapi$), as given by \citet{2016A&A...588A..87V}. The technique is based on the stretching of the frequencies in the oscillation spectrum \citep{2015A&A...584A..50M} in order to obtain a mixed-mode pattern spaced regularly in period, corresponding to $\deltapi$.

The radial ($\ell = 0$) and quadrupole ($\ell = 2$) modes are first identified using the universal red giant oscillation pattern as described by \citet{2011A&A...525L...9M} then suppressed from the spectra. Then, the modification of the spectrum frequencies is applied and it follows that the regularity of the g-mode pattern can be measured using a Fourier Transform. This allows the measurement of the $\deltapi$ parameter and, therefore, permit the determination of the stellar evolutionary status.  

\subsection{Second method ($\Mdeux$): \citet{2019A&A...622A..76M}}

This second method is based on the measurement of the $\Dnu$ signal present on the autocorrelation of the star's time-series. 
The signature of the pressure mode pattern can be retrieved by performing a Fourier Transform of the filtered Fourier spectrum (EACF) around the observed oscillations. This corresponds in this case to an autocorrelation \citep{2009A&A...508..877M}. The obtained EACF signal can therefore be measured and it appears that this signal is notably lower by a factor of $2.5$ for RC stars compared to RGB stars at same $\Dnu$ values. This difference is partly, if not totally, due to the fact that RC pressure modes have shorter lifetimes than RGB ones \citep{2018A&A...616A..94V}. This behavior can therefore be used to determine the evolutionary status of red giants. 

Because this method is not using the mixed modes pattern or the global seismic parameters, \citet{2019A&A...622A..76M} has also used it to differentiate AGB and RGB stars. 
At first approximation, for the construction of the determined evolutionary status sample, the stars labeled as AGB by this method were included in the RGB sample.

\subsection{Third method ($\Mtrois$): \citet{2017MNRAS.469.4578H,2018MNRAS.476.3233H}}

Method $3$ is a machine learning technique based on 1D convolutional neural networks. For each spectrum, the frequency range where the oscillations are present is folded around $3\,\Dnu \pm \nu_{\mathrm{max}}$. The program is then trained on labelled data in order to allow it to recognize the features corresponding to RGB or RC power spectra. The clear difference in the mixed mode pattern between RC and RGB stars is likely the main feature that allows that determination \citep{2017MNRAS.469.4578H}. Finally, the program is used on non-classified red giants. The success rate of the classification was found to be very high, even for stars with low signal-to-noise spectra \citep{2018MNRAS.476.3233H}.

\subsection{Fourth method ($\Mquatre$): \citet{2017MNRAS.466.3344E}}

This method is based on the measurement of the mixed mode period spacing ($\dP$) in the region of the acoustic spectrum where the oscillations are visible. The location of the $\ell = 1$ mixed modes in the acoustic spectrum is determined using the Universal Pattern \citep{2011A&A...525L...9M}. This information is used to exclude the even $\ell$ modes from the analysis. A threshold level, based on the signal-to-noise in the spectrum, is applied to the regions around the $\ell=1$ mixed modes  and the frequency of all the data points above this threshold is noted. Now working in the period domain, a  histogram is formed from the difference in period, between each feature and all the other features within about $0.3\Dnu$.
Because the $\deltapi$ parameter (and hence $\dP$) is much larger for RC stars than it is for RGB stars, the location of the peak in the histogram allows the classification of the evolutionary status.
The evolutionary status is therefore determined following the $\dP$ and $\numax$ values. More information on the precise definitions of $\dP$ and $\numax$ are given in \citet{2017MNRAS.466.3344E}. This method was proven to be effective in its classification for spectra with resolved and not depressed mixed-modes.

\subsection{Fifth method ($\Mcinq$): \citet{2020MNRAS.497.4843K}}

This method uses a machine learning technique and is based on the light-curve characteristics, in contrast to the previous ones which use the power spectrum. The algorithm corresponds to a supervised classification algorithm and is based on the principle that several light-curve characteristics (variance, magnitude, signal coherency, normalized number of zero crossing, median of time-series first differences) are slightly different as a function of the stellar evolutionary status. For each star, these characteristics are measured. The program is, then, trained to classified RGB and RC stars based on already classified data. After that, the program is applied on unclassified red giants. If there is less than a $33\%$ chance that a star is RC (respectively, RGB) it will be classified as RGB (respectively RC). According to \citet{2020MNRAS.497.4843K}, the false positive rate for this technique is lower than $8\%$.

\subsection{Sixth method ($\Msix$): \citet{2012A&A...541A..51K}}

This last method is based on the precise pressure mode pattern. This technique locates the radial ($\ell = 0$) modes in each spectrum and deduces from those frequencies the value of the phase shift $\Eps$ of the central radial mode, $\Eps$ corresponding to the offset in the linear asymptotic fit to the acoustic modes.
At given $\Dnu$, this parameter is significantly different for RGB stars compared to the RC stars \citep{2012A&A...541A..51K}. 
This behavior can be related to deviations from the asymptotic relations due to sharp changes in the internal structure of stars \citep{2015A&A...579A..84V}. The method was proven to be efficient even at low signal-to-noise \citep{2012A&A...541A..51K,2019arXiv190609428K,2021A&A...650A.115D}. However, there is some overlapping between $\Eps$ values at the same $\Dnu$ for RGB and RC stars \citep{2012A&A...541A..51K}, which can lower the confidence of the classification.

Because this method uses only the pressure mode pattern it can be extended to the distinction between AGB and RGB stars.
Here, as for $\Mdeux$, the stars labeled as AGB by this method were included in the RGB sample as a first approximation for the construction of the determined evolutionary status sample.

\section{Forging a seismic evolutionary status consensus sample \label{seismic_consensus}}

In this section, we will describe the different results that were obtained for each of the previously described methods and explain how we selected the final sample containing consensus values on the star's evolutionary stages.  

\subsection{Result description}

\begin{table*}[h]
\centering
\begin{tabular}{lccccccc}

Seismic evolutionary status &  $\MunA$ & $\MunB$ & $\Mdeux$ & $\Mtrois$ & $\Mquatre$ & $\Mcinq$ & $\Msix$\\
\hline
Red giant $\Kepler$ legacy sample & $7756$ & $8102$ & $24404$ & $24716$ & $16580$ & $21941$ & $6179$ \\
-RC & $4603$ & $4929$ & $10218$ & $8405$ & $6726$ & $8590$ & $2569$ \\
-RGB/AGB & $3153$ & $3173$ & $14186$ & $16311$ & $9854$ & $13351$ & $3610$ \\
\hline
APOKASC-3 sample & $4801$ & $5017$ & $14237$ & $14302$ & $10116$ & $14138$ & $6151$ \\
-RC & $3233$ & $3409$ & $5753$ & $5207$ & $4402$ & $6088$ & $2561$ \\
-RGB/AGB & $1568$ & $1608$ & $8484$ & $9095$ & $5711$ & $8050$ & $3590$ \\
\hline
\end{tabular}
\caption{Number of evolutionary status results given by the different seismic methods for the red giant $\Kepler$ legacy sample (Garcia et al., in preparation) and the APOKASC-3 \citep{2024arXiv241000102P} stars. \label{Table_seismicresults}}
\centering
\end{table*}

The results from the different techniques are shown in table \ref{Table_seismicresults}. Concerning Method $1$, two different codes were used as described in \citet{2016A&A...588A..87V}. In the rest of the text, the different techniques will be referred as the results from Method $\MunA$ and $\MunB$.

We can see that the machine-learning techniques ($\Mtrois$, $\Mcinq$) and the method $\Mdeux$ provide the highest number of results. This is understandable, because these automated techniques are based, respectively, on the shape of the spectra, the light-curves parameters and the pressure mode pattern. Those characteristics can indeed be measured or recognized in an easier way than mixed-modes, therefore allowing a determination of the evolutionary status for a larger number of stars. On the contrary, method M$1$ and $\Mquatre$ give a lower number of results, mainly for RC and RGB stars with $\numax$ $>$ $20\mu$Hz because mixed modes disappear for high-luminosity red giants due to lower coupling between the resonant cavity of the two types of modes \citep[see e.g.:][]{2014A&A...572A..11G}. The only exception concerns method $\Msix$, which is also based on the analysis of the pressure mode pattern but gives a result for a limited number of stars. This can be understood by the fact that this technique needs a precise identification of the modes, which is not easy to realize with a high confidence for a large number of stars. The sample of analyzed stars was therefore restricted for this technique.
\newline

The automated techniques that gave the most results ($\Mdeux$, $\Mtrois$ and $\Mcinq$) also have a few stars they identified as RC in their sample and that possess spectroscopic characteristics ($\Teff$ and $\logg$) that correspond to the lower part of the RGB. These probable misidentifications are all located near the Nyquist frequency for $\Kepler$ targets, which can explain the confusion for automated techniques because the spectra become difficult to analyze at those frequencies. We will address the case of these stars later in the paper.

\subsection{Merging classification results \label{Merging_classification}}

For each method, the stars were classified into three categories: RGB, RC or undetermined. We used the comparison between the different methods to improve the robustness of the evolutionary status determination as was done by \citet{2019MNRAS.489.4641E}. Putting together the classification provided from those methods was made following five different scenarios: 
\newline

The first scenario happens when at least three methods agrees with each other on the evolutionary status classification and no technique disagrees. In this scenario, some methods may not provide any results, therefore considering the star's evolutionary status as unclassified. This is not considered as a disagreement between the methods. In that situation, the consensus on the classification is clear and the star can be considered as a RGB or RC star following what the methods agreed on. The classification is considered as being robust.

The second and third scenarios arise when at least three methods agrees with each other but another one disagrees, the other ones giving an undetermined status classification. In that situation, we looked at which method disagrees with the consensus because some methods are probably less reliable than others. For that work, we considered the methods that based their classification on the mixed modes analysis as the most reliable because it is principally this feature present in the oscillation spectra that allows a distinction between RC and RGB stars. The other methods like the ones presented in \citet{2012A&A...541A..51K} ($\Msix$) and \citet{2019A&A...622A..76M} ($\Mdeux$) present an overlap between the two populations for the classification criterion they considered. $\Mcinq$ \citep{2020MNRAS.497.4843K} has also a lower success rate, around $92$ $\%$, than the methods that use mixed modes. Therefore, when the only disagreement comes from the methods $\Mdeux$, $\Mcinq$ or $\Msix$, we will follow the classification from the other methods. However, when the disagreement comes from methods $\MunA$, $\MunB$, $\Mtrois$ or $\Mquatre$, we classify the star as having an uncertain evolutionary status.

A fourth scenario occurs when two methods or more are giving an evolutionary status determination that contradicts two other methods or more. In that situation, the star is classified as having a conflict in the evolutionary status classification. We will discuss these stars in more detail in part \ref{Sec_Conflict}.

The last scenario corresponds to when two or fewer methods are giving a classification result. In that situation, we considered that a consensus is not reached on the evolutionary status. Therefore, the star is classified as having an undetermined evolutionary status.
\newline

Below is a summary of the classification process for the evolutionary status (here noted EV) as a function of the number of agreements (A) and disagreements (D) between the different methods.  

\begin{tikzcd}[cramped, sep=small]
\geq 3A , \,0D \arrow[dr] \\
\geq 3A , \,1D\,(\Mdeux,\Mcinq,\Msix) \arrow[r] & robust\,EV \\
\geq 3A , \,1D\,(\MunA,\MunB,\Mtrois,\Mquatre) \arrow[r] & uncertain\,EV \\
\geq 2A , \, \geq 2D \arrow[r] & conflict \\
\leq 2A , \, \leq 1D \arrow[r] & undetermined\,EV \\                       
\end{tikzcd}

\subsection{Classification results and discussion}

\begin{figure}                 % Insertion d'une figure = objet flottant
  \includegraphics[width=9cm]{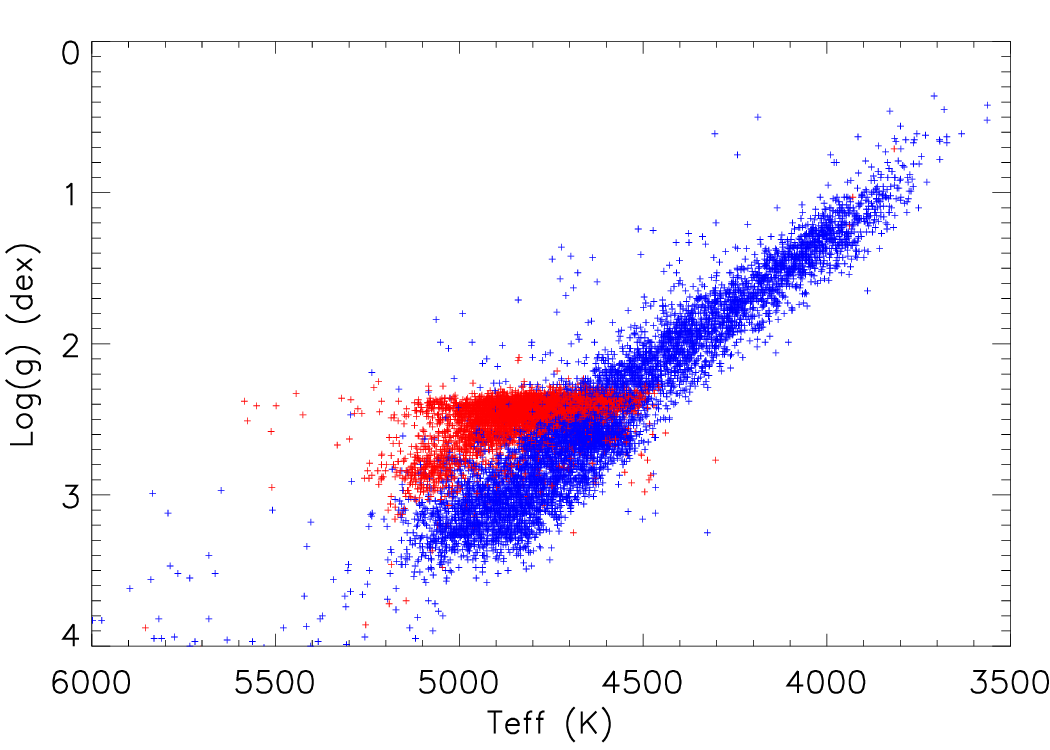}
  \caption{Stellar surface gravity as a function of the effective temperature $\Teff$ for the stars classified as having a robust evolutionary status. The red stars represent the RC and the blue stars represent the RGB or AGB stars for high-luminosity targets.\label{fig:agreement_evolution}}
\end{figure}

\begin{table*}[h]
\centering
\begin{tabular}{ccccc}

%add in the discussion the number of quarters so we know which light-curves are high-quality.

KIC number &  $\numax$ ($\mu$Hz) & $\Dnu$ ($\mu$Hz) & classification status & evolutionary status \\
\hline
 757076  &   261.7   &  18.79  &  1  &  1 \\
 757137  &    30.0   &   3.39  &  1  &  1
\\ 
 892010  &    18.0   &   2.44  &  1  &  1 \\
 892107  &   274.5   &  17.64  &  1  &  1 \\
\hline
\end{tabular}
\caption{First lines of a joined file summarizing the evolutionary status determination. The different columns correspond respectively to the KIC number, preliminary $\numax$, preliminary $\Dnu$, the classification result ($0$ for undetermined, $1$ for robust) and the evolutionary status ($1$ for RGB/AGB, $2$ for RC, $-1$ for no classifications). \label{Table_example_evolutionary_status_results}}
\centering
\end{table*}

From this classification, we obtained an evolutionary status for  $18784$ stars ($11387$ RGB and $7397$ RC). Among them, we have APOGEE data for $11516$ stars ($6732$ RGB and $4784$ RC). The results are displayed in Figure (\ref{fig:agreement_evolution}) for APOGEE data and the complete results are available at the CDS. The first lines of the data file are given as an example in Table \ref{Table_example_evolutionary_status_results}. The RC appears in a well defined compact line with $2<\logg<3$. However, a few stars, identified by several methods as RC stars, appear to have particularly high and low surface gravity values. We will discuss about those stars in more detail in Section \ref{Clump_limit}. The RGB (or AGB) are distributed along all the red giant branch except at high and low surface gravity. The low red giant branch is not visible because the detection of seismic oscillations below the Nyquist frequency (corresponding to approximately $283.0\mu$Hz for $\Kepler$ long-cadence data) is very difficult and will greatly affect possible consensus on the evolutionary status. The tip of the RGB or AGB is also not visible: the detection of the oscillation for highly luminous RGB and AGB becomes very difficult because of the length of the pulsations and the small number of visible modes \citep[e.g.][]{2013A&A...559A.137M,2014ApJ...788L..10S}. The length of the observation does not ensure a high enough spectral resolution to obtain an estimation of the seismic parameters, therefore hampering a possible consensus on the evolutionary status for those stars. We note that, above the RC, fewer RGB or AGB stars have an identified evolutionary status. The cause of this comes from the lower number of methods retrieving an answer for these red giants.  

We also note that a few stars appears to possess values of spectroscopic parameters far away from what we expect for red giants. We have a few hundred with spectroscopic parameters closer to main-sequence stars than to red giants. This behavior corresponds in part to the fact that the spectroscopic measurement can be biased by the presence of background stars with different spectroscopic parameters \citep{2012ApJS..199...30P} and, in part, from the selection of the red giant $\Kepler$ legacy sample that contains a few targets wrongly identified as giants because their aperture are polluted by nearby red giants (Garcia et al. in preparation). These targets will be checked and modified in the published version of the red giant $\Kepler$ legacy sample. Four stars with very low spectroscopic surface gravity and identified as RC by seismology are also present in the final sample. Those targets are identified as likely belonging to background stars because their spectroscopic $\logg$ values deviates more than $2$ $\sigma$ from the seismic ones.
\newline

\begin{figure}                 % Insertion d'une figure = objet flottant
  \includegraphics[width=9cm]{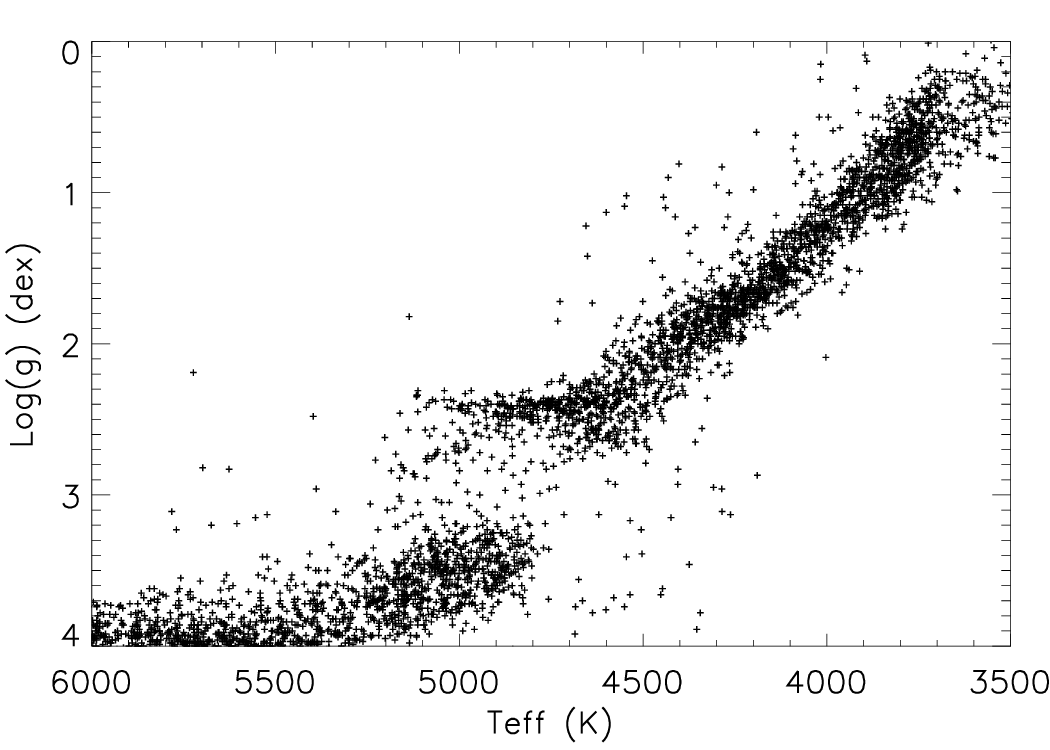}
  \includegraphics[width=9cm]{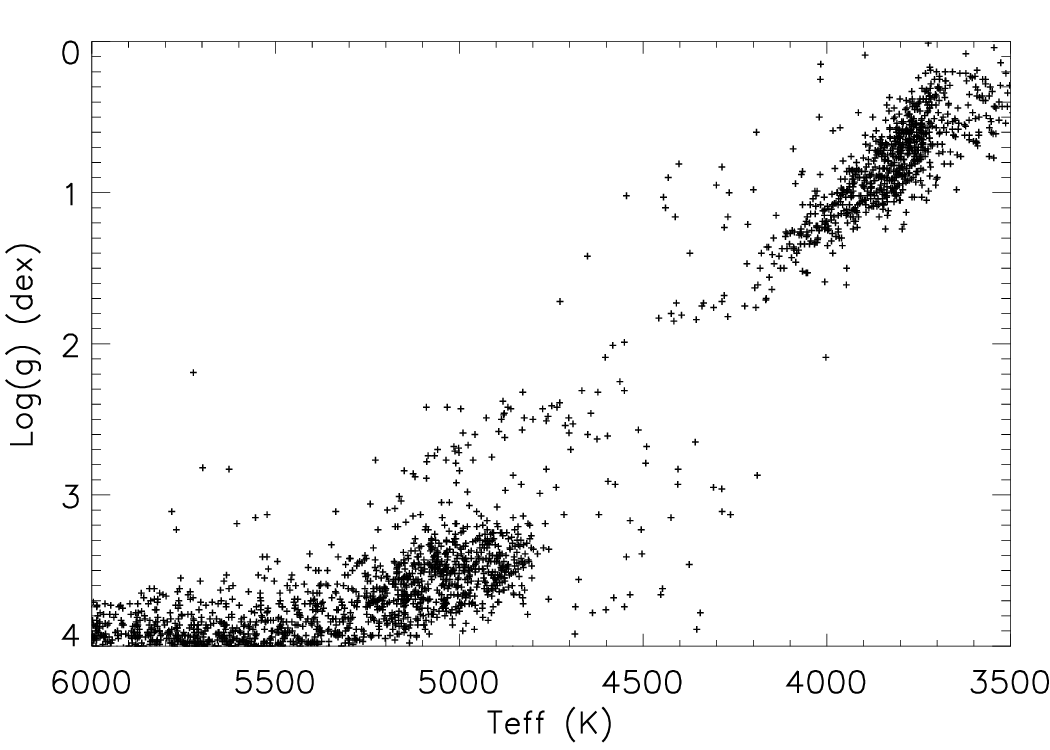}
  \caption{Spectroscopic stellar surface gravity as a function of the effective temperature $\Teff$ for the stars having an undetermined evolutionary status (Top: $5276$ APOKASC-3 targets) and a duration of observation higher than $4$ months (Bottom: $3670$ APOKASC-3 targets). \label{fig:no_data}}
\end{figure}

The stars with undetermined evolutionary status due to too few results from the different methods are shown in a Kiel diagram in Figure \ref{fig:no_data}. There are $5276$ such APOKASC-3 stars ($9211$ for the full red giant legacy sample). Most of those correspond to stars with very low or very high surface gravity, particularly stars with a $\numax$ value near the Nyquist frequency. This is consistent with the difficulty to obtain a reliable seismic measurement for these stars. We note particularly that many of these stars have spectroscopic parameters corresponding to main-sequence or subgiants rather than giants. Those characteristics correspond certainly to blended sources in the $\Kepler$ pixels: stars with different spectroscopic parameters blended in the same KIC number. The seismic signature we perceive is thus related to the background red giant star and not the main spectroscopic target. Those faint red giants are particularly difficult to analyze, therefore their evolutionary status will not be determined by the different automated techniques, consistent with the results we obtained. In the bottom part of Figure \ref{fig:no_data}, we can see that most of those targets correspond to stars with long light-curves. It is also clear from Figure \ref{fig:no_data} that we can detect and classify the vast majority of giants with intermediate surface gravities. It demonstrates that even for long light-curves, the classification for low-$\logg$ stars is not always easy, probably due to the disappearance of the mixed-modes and the decrease of $\deltapi$ towards the RGB values. A few stars with spectroscopic parameters consistent with RC are also present in the top part of Figure \ref{fig:no_data}. Those red giants do not appear in the bottom part of the Figure, therefore showing they correspond to stars with short light-curves. Therefore, these targets have low signal-to-noise spectra and more complex spectral windows, which explain why the determination of their evolutionary status is difficult and no consensus were reached from the different methods.
\newline

\subsection{Conflicted evolutionary status \label{Sec_Conflict}}

Among the stars that were not classified as RC or RGB, only $452$ of APOKASC-3 red giants have an evolutionary status that correspond to the fourth scenario when a clear conflict arises between the different methods ($650$ in the total red giant legacy sample). In order to understand where this conflict arises we performed a thorough analysis of the evolutionary status of this sample. We checked individually the $\deltapi$ values of those stars following the method of \citet{2014A&A...572L...5M} and \citet{2016A&A...588A..87V}. We confirmed an unambiguous measurement for $109$ stars in this sample. The results are displayed in Table \Ref{Table_conflict}. It is important to note that very few ($< 10$) of these possess short light-curves ($< 4$ months). Therefore, a determination of the evolutionary status should be possible in those targets.

\begin{table*}[h]
\centering
\begin{tabular}{cccccccc}

%add in the discussion the number of quarters so we know which light-curves are high-quality.

Conflict stars &  $\MunA$ & $\MunB$ & $\Mdeux$ & $\Mtrois$ & $\Mquatre$ & $\Mcinq$ & $\Msix$\\
\hline
Total & $118$ & $212$ & $451$ & $451$ & $420$ & $449$ & $243$ \\
$\%$ of robust EV & $2.45$ & $4.23$ & $3.17$ & $3.15$ & $4.15$ & $3.18$ & $3.95$ \\
With measured $\deltapi$ & $56$ & $73$ & $109$ & $108$ & $103$ & $108$ & $70$ \\
With incorrect EV & $15$ & $12$ & $67$ & $40$ & $10$ & $57$ & $44$ \\
$\%$ correct EV & $73.2$ & $83.6$ & $38.5$ & $63.0$ & $90.3$ & $49.5$ & $37.1$ \\
\hline
\end{tabular}
\caption{Stars for which a conflict is present between the different evolutionary status that was given by the different seismic methods. The first line show the total number of stars with conflict for each method. The percentage of conflict stars compared to the total red giant $\Kepler$ legacy sample of robust evolution status is given on the second line. The third line correspond to the number of conflict stars with measured and checked $\deltapi$. The fourth line gives the number of incorrect evolutionary status for the stars with checked $\deltapi$. The fifth line show the percentage of correct evolutionary status for the stars with checked $\deltapi$. \label{Table_conflict}}
\centering
\end{table*}

Do you mean "The fraction of stars with disagreements between methods on the evolutionary status is $2.5$-$4.2\%$, thus showing the high level of agreement between the methods. We can also see that the correct evolutionary status for conflict stars, following $\deltapi$ measurement, is deduced correctly for more than $70$ $\%$ of red giants for $\MunA$, $\MunB$ and $\Mquatre$. This is consistent with the fact that those techniques based on the mixed modes are more reliable and we confirm our selection of those techniques as more reliable when a conflict is arising. We can note also the good reliability of $\Mtrois$, which tends to agree with the verified $\deltapi$ values for more than half of the case. This result can be understood because this method is based on the shape of the oscillation spectra around the pure pressure $\ell =1$ modes, therefore on the mixed-mode pattern global appearance. $\Mdeux$, $\Mcinq$ and $\Msix$ show less reliability. Because those techniques are based on secondary indicators or parameters (pressure-mode pattern, light-curve analysis, amplitude of the oscillation envelope) for which the two evolutionary status overlap, this behavior is not surprising. These results highlights the reliability of the mixed-mode based methods when a conflict arises. These results confirm our overall classification approach (Section \ref{Merging_classification}), where we adopted the consensus value if a single method disagreed.

For the rest of the analysis, we will now focus on the susbsample containing the APOKASC-3 data in the red giant $\Kepler$ legacy sample, therefore allowing to perform a more detailed analysis of this sample and its evolutionary status determination.

\section{Distinguishing RC and RGB stars with spectroscopic data}\label{Section:Spectro_Determination_technique}

The APOGEE survey uses a multi-layered approach for measuring stellar properties. In the first set of stages, an automated pipeline is used to infer trial values for key observables ($\Teff$, $\logg$, the ratios of metals, carbon and nitrogen: respectively [C/M] and [N/M], and the alpha to iron ratio: [$\alpha$/Fe]). These trial values are then placed on an absolute system in a post-processing calibration step. The assignment of a spectroscopic evolutionary status is an important ingredient for inferring spectroscopic surface gravities because of a still poorly-understood phenomenon: the difference between the trial $\logg$ (here $\ginit$) and calibrated $\logg$ (here $\gcal$) is observed to be a function of evolutionary status.

Spectroscopic evolutionary status take advantage of the fact that the RC exists only in a narrow $\logg$ range, and that RC stars are hotter than first ascent RGB stars on average. The $\Teff$ difference between them, of order $200$K, is found to have a weak dependence with mass, metallicity, and surface gravity, although the absolute effective temperatures of both populations are sensitive to all three. The APOGEE survey therefore defines a reference surface temperature ($\Tref$) as a function of $\logg$, metallicity, and carbon-to-nitrogen: [C/N] (a spectroscopic proxy for mass),

\begin{multline}
\label{Eq:Teff_ref2}
    \Tref = \\ 3032.8 + 552.6\mathrm{log}(\ginit) -488.9\mathrm{[M/H]} -357.1\mathrm{[C/N]}
\end{multline}

Stars with $3.5$ $>$ log($\ginit$) $>$ $2.38$ and $\Teff$ $>$ $\Tref$ were classified as RC stars in DR$16$ \citep{2020AJ....160..120J}, and they were assigned a surface gravity

\begin{equation}
\label{Eq:gref}
    \mathrm{log}(\gcal) = \mathrm{log}(\ginit) + 4.532 - 3.222\mathrm{log}(\ginit) + 0.528 (\mathrm{log}(\ginit))^2.
\end{equation}

All other stars with log($\ginit$) $<$ $3.5$ had the RGB correction applied,

\begin{multline}
\label{Eq:gref2}
\mathrm{log}(\gcal) = \mathrm{log}(\ginit) - (-0.441 + 0.759 \mathrm{log}(\ginit) - \\ 0.267 (\mathrm{log}(\ginit))^2 + 0.028 (\mathrm{log}(\ginit))^3 + 0.135[M/H]) ,
\end{multline}

\noindent
while dwarfs with log($\ginit$) $>$ $4$ had the following correction applied:

\begin{equation}
\label{Eq:gref3}
\mathrm{log}(\gcal) = \mathrm{log}(\ginit) - (-0.947 + 1.886 \ 10^{-4} \Teff{}_{\!,{\rm spec}} + 0.410\mathrm{[M/H]})
\end{equation}

\begin{figure}                 % Insertion d'une figure = objet flottant
  \includegraphics[width=9cm]{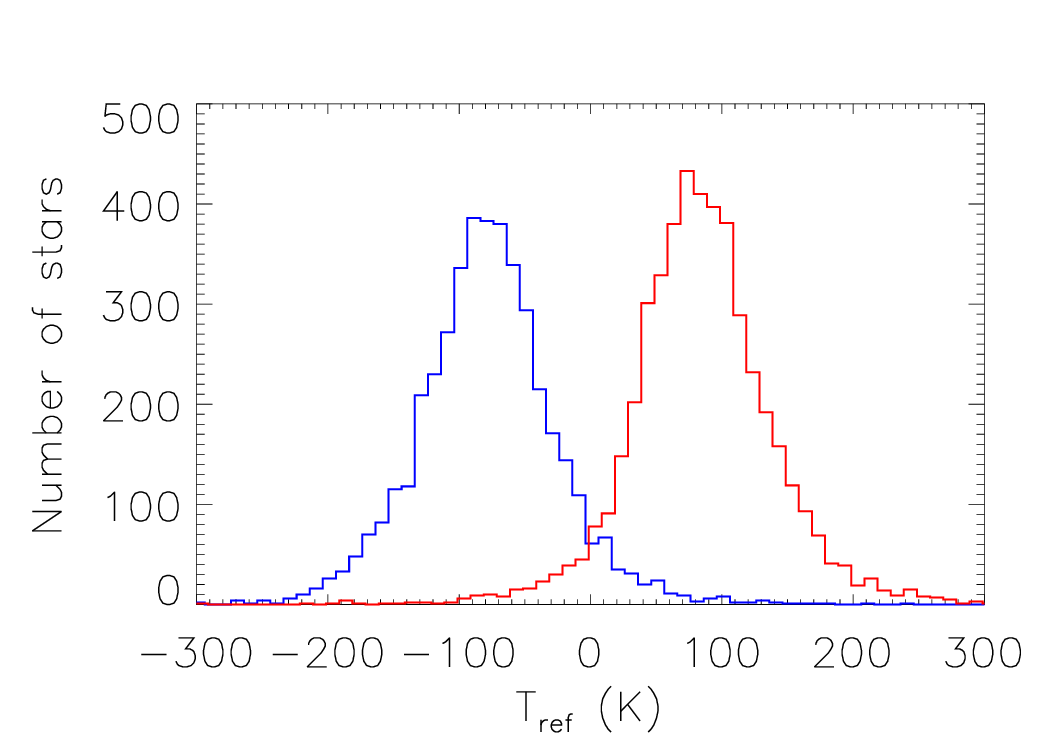}
  \caption{Distribution of stars as a function of $\Tref$. The blue and red histograms correspond, respectively to RGB/AGB and RC stars as determined by seismology following Section \ref{seismic_consensus}. \label{fig:Distribution_DeltaT}}
\end{figure}
 
Subgiants with $3.5$ $<$ $\logg$ $<$ $4.0$ had a weighted average of the dwarf and RGB values assigned. These surface gravity assignments are important for what follows, because stars with an erroneous spectroscopic evolutionary status will have the wrong correction applied, which moves them away from their true position in the surface gravity-temperature plane. Figure \ref{fig:Distribution_DeltaT} shows the stars asteroseismically classified as RGB (blue) and RC (red) clearly separated in the $\Tref$ plane. The large majority of stars fall naturally on the positive side or the negative side of $\Tref$, but a significant subset do not. Those red giants will likely be spectroscopically classified incorrectly relative to the seismic classification. We fitted a Gaussian function over the two evolutionary status samples and found that the width at half maximum of each sample correspond to $44.04$ $\pm$ $0.43$K for RGB and $43.28$ $\pm$ $0.47$K for RC. Because the position at the center of the peak of each sample is $-79.46$ $\pm$ $0.43$K for RGB and $82.31$ $\pm$ $0.47$K for RC, the two evolutionary status will overlap at less than two times the width at half-maximum of their $\Tref$ distribution. The overlap between the two samples, which will produce a misidentification of the evolutionary status, will therefore correspond to more than $5\%$ of the total sample. Here, it corresponds to $5.6\%$ of the stars in our sample. These cases will be discussed in Section \ref{Divergent_classification}.

\section{Seismic and spectroscopic evolutionary status comparison}

In the APOKASC-3 sample, the seismic evolutionary status is determined for $11371$ stars ($4755$ identified as RC and $6616$ identified as RGB or AGB) out of $15464$ red giants, while the spectroscopic evolutionary status was determined for $15232$ stars. Because the seismic and spectroscopic evolutionary status are not obtained uniformly for the whole sample, we will focus in this section on the $11297$ stars for which both seismic and spectroscopic classification give an evolutionary status classification.

\subsection{Agreement between seismic and spectroscopic classification}

In this sample, an agreement between the two evolutionary status determination techniques was found for $10780$ red giants ($4530$ RC and $6250$ RGB or AGB stars), thus $94.80\%$ of the stars where both evolutionary status determination techniques gave a result. This highlights the agreement between the different methods and the accurate calibration of the spectroscopic evolutionary status determination technique as stated by \citet{2019MNRAS.489.4641E}. On the contrary, $517$ red giants have divergent evolutionary status results. A discussion about these stars is present in Section (\ref{Divergent_classification}).

\begin{figure}                 % Insertion d'une figure = objet flottant
  \includegraphics[width=9cm]{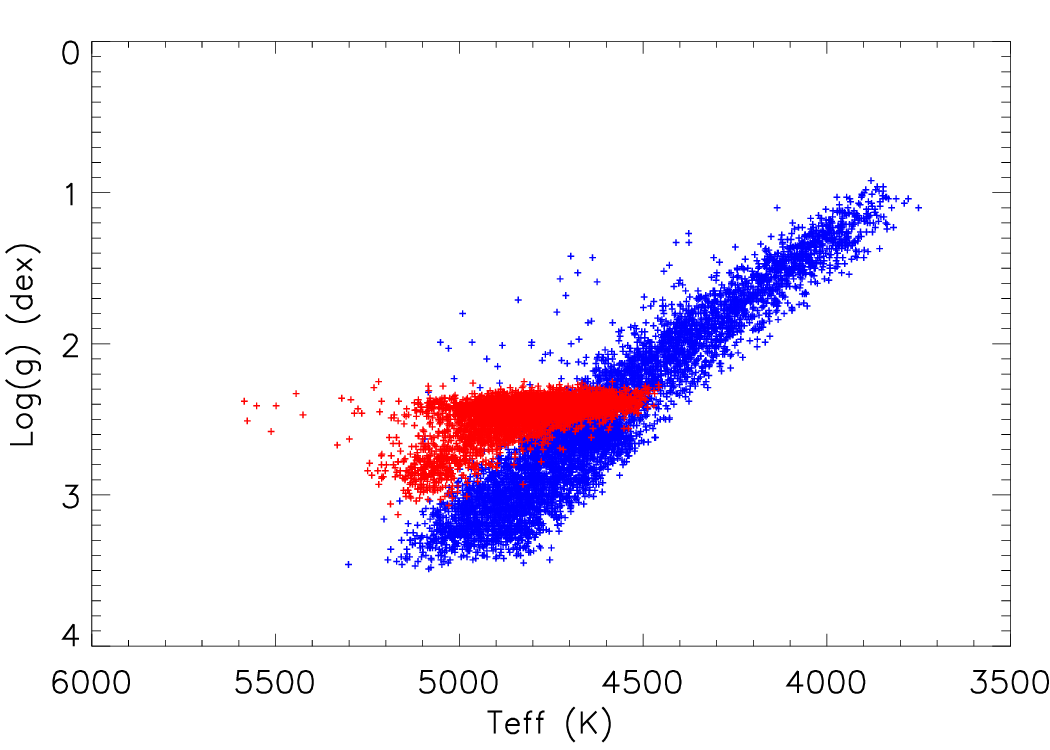}
  \caption{Stellar surface gravity as a function of the effective temperature $\Teff$ for the stars showing agreement between the spectroscopic and seismic evolutionary status classification. The red stars represent the RC and the blue stars represent the RGB or AGB stars for high-luminosity objects. \label{fig:Compare_spectro_sismo}}
\end{figure}

The results, for the stars where the evolutionary states agree, can be seen on Figure \ref{fig:Compare_spectro_sismo}. The RC is well defined and separated from the RGB and AGB stars. The secondary clump appears also very clearly and no sample of stars appears to deviate significantly from what is expected for clump and RGB stars.

\subsection{Stars with a divergent evolutionary status classification}\label{Divergent_classification}

\begin{figure}                 % Insertion d'une figure = objet flottant
  \includegraphics[width=9cm]{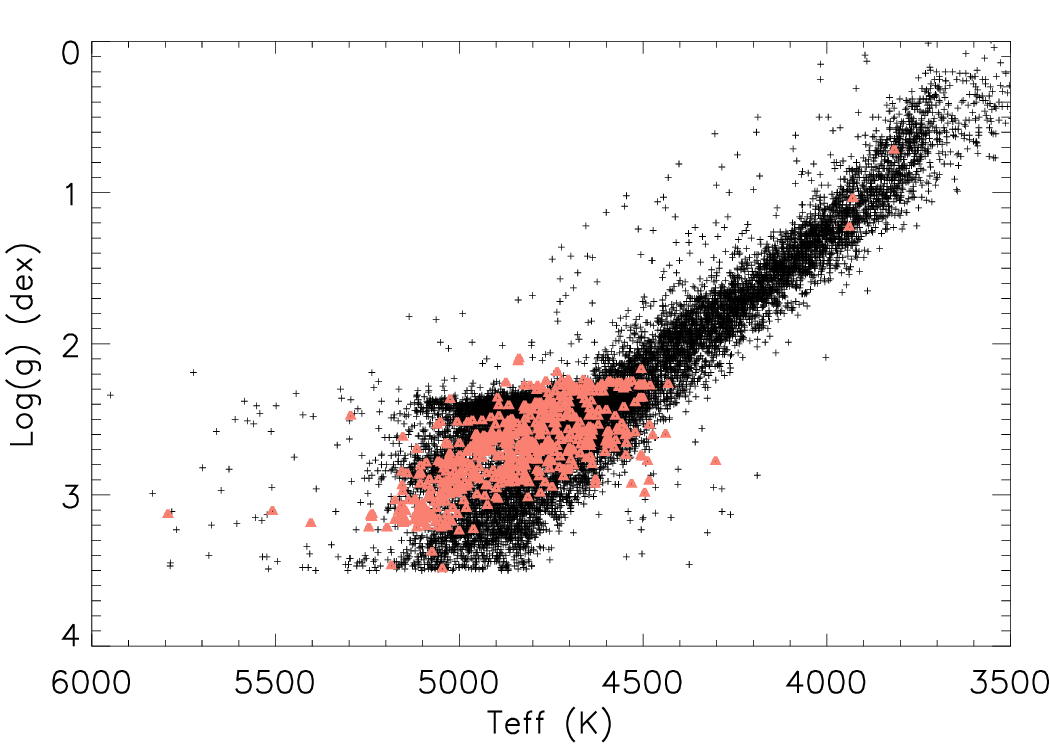}
  \includegraphics[width=9cm]{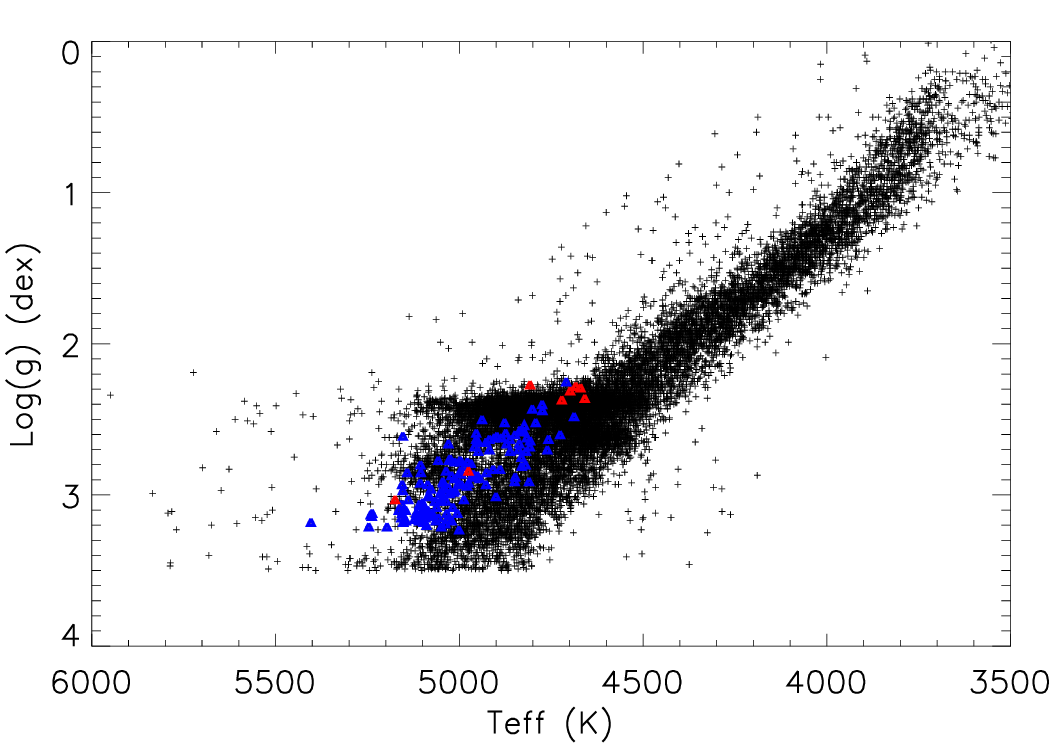}
  \includegraphics[width=9cm]{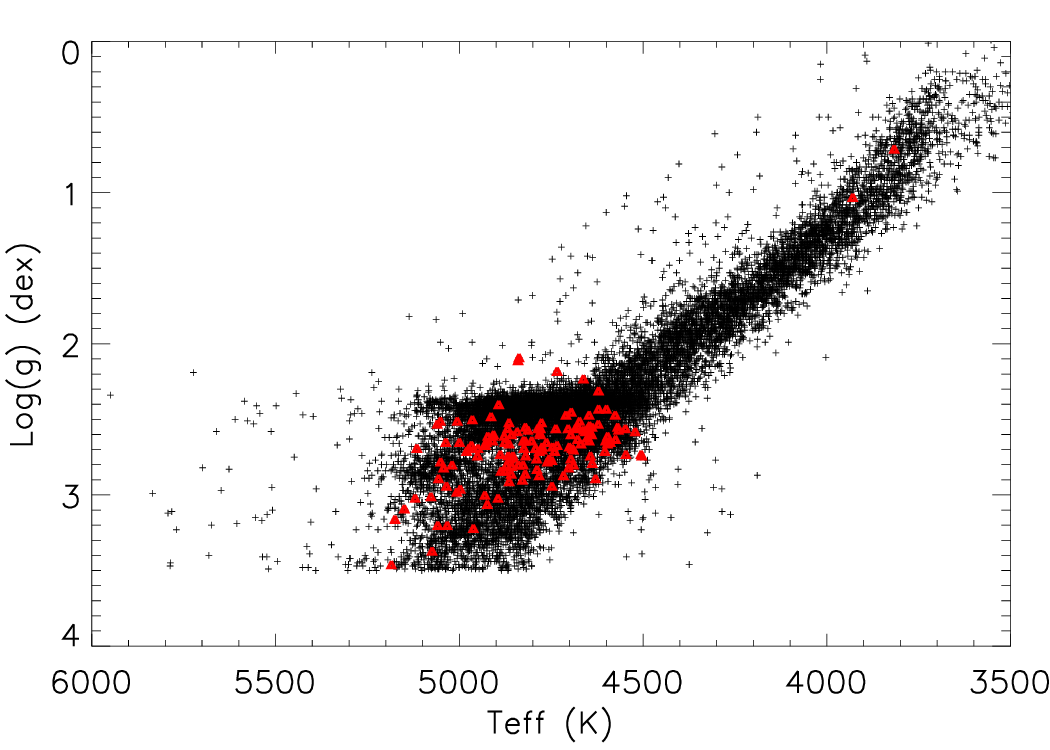}
  \caption{Stellar surface gravity as a function of the effective temperature $\Teff$ for the stars with an agreement between the spectroscopic and seismic evolutionary status classification (in black). \textit{Top}: The Salmon triangles show the stars with divergent spectroscopic and seismic evolutionary status. \textit{Center}: The colored triangles represent the stars identified as RGB by seismology and RC by spectroscopy for which the evolutionary status was confirmed as RGB (blue) or RC (red). \textit{Bottom}: Same as center but with the red triangles corresponding to confirmed RC stars identified as RC by seismology and RGB by spectroscopy. \label{fig:Disagreement_all}}
\end{figure}

Divergent evolutionary status between spectroscopic and seismic techniques are present for $517$ stars. $215$ stars are identified as RC by seismology and as RGB by spectroscopy. On the contrary, $302$ stars are recognized as RGB by seismology and as RC by spectroscopy. Their positions in the HR diagram are shown on the top part of Figure \ref{fig:Disagreement_all}. These stars are concentrated near the RC because this is the domain where the two populations overlap. The only exceptions concern $4$ stars for which the low spectroscopic surface gravity (around $1$) does not correspond to the seismic ones which are way higher (between $2.39$ and $2.5$ for those red giants). These inconsistencies can explain the misidentification of those stars as belonging to the RGB or AGB by the spectroscopic classification. A more thorough discussion on the discrepancies between spectroscopic and seismic $\logg$ is present in Section \ref{Clump_limit}.
\newline

To confirm the evolutionary status of each star with conflict between the seismic and spectroscopic classifications, a specific check of the mixed modes was performed. The evolutionary status was considered to be certain when the measurement of $\deltapi$ following the method of \citet{2014A&A...572L...5M} and \citet{2016A&A...588A..87V} was possible. The stars with $\Dnu$ $>$ $10$ $\mu$Hz were also automatically identified as RGB following the results of \citet{2013ApJ...765L..41S} and \citet{2014A&A...572L...5M}. The center and bottom part of Figure \ref{fig:Disagreement_all} show the precise results for the stars with confirmed evolutionary status.
\newline

The stars identified as RC by seismology and RGB by spectroscopy was confirmed to belong to the RC for $143$ stars out of $215$. None were identified as RGB stars. We can see on the bottom part of Figure \ref{fig:Disagreement_all} that a significant part of these stars correspond to secondary clump stars and the high surface gravity RC stars previously identified. Because of their high $\logg$ values, these stars have higher chances to be classified as RGB by the spectroscopic classification method. A further discussion about these stars is present in Section \ref{Clump_limit}.

The stars identified as RGB by seismology and RC by spectroscopy were confirmed to belong to the RGB for $145$ red giants and to the RC for $8$ out of $302$. The stars identified as true RGB correspond mostly to high-mass stars ($83$ stars out of $145$ have a mass larger than $1.5\Msol$) that are spectroscopically confused as RC because of their high $\Teff$ and mass, putting them closer to secondary-clump in the $\Teff$-$\logg$ space. In this configuration, The [C/N] ratio becomes insensitive to mass for higher mass stars \citep{2024MNRAS.530..149R}, so this is not surprising. The stars identified as true RC have, except for two of them, low values of $\deltapi$ and $\Dnu$, meaning they are on their way towards the AGB phase \citep{2013ApJ...765L..41S,2014A&A...572L...5M}. Their lower seismic parameters certainly explain why several seismic methods identified those stars as belonging to the RGB or the AGB. However, because the $\deltapi$ of these stars is still much higher than those of RGB at the same $\Dnu$, they are still in the helium core burning phase.
\newline

\section{Evolutionary status of specific populations and RC lower limit}\label{Clump_limit}

Figure \ref{fig:agreement_evolution}, \ref{fig:Compare_spectro_sismo} and the bottom part of \ref{fig:Disagreement_all} show several RC stars with surface gravities much higher than expected for true RC stars. We checked that the spectroscopic surface gravity is consistent with the asteroseismic one. The seismic surface gravity (g) was obtained through the scaling relation
\citep[e.g.][]{1995A&A...293...87K}:

\begin{equation}
\label{Eq:echelle_law}
    g = \log\left(\frac{\numax}{3076} \sqrt{\frac{\Teff}{5772}}\right) + 4.437.
\end{equation}

\begin{figure}                 % Insertion d'une figure = objet flottant
  \includegraphics[width=9cm]{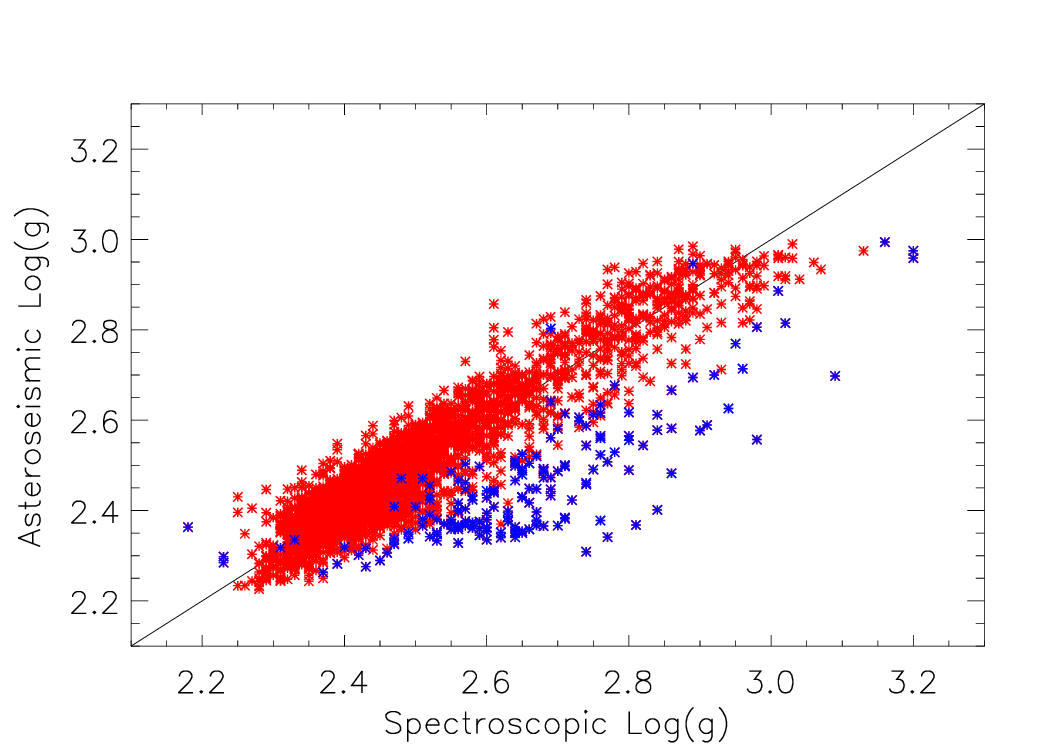}
  \includegraphics[width=9cm]{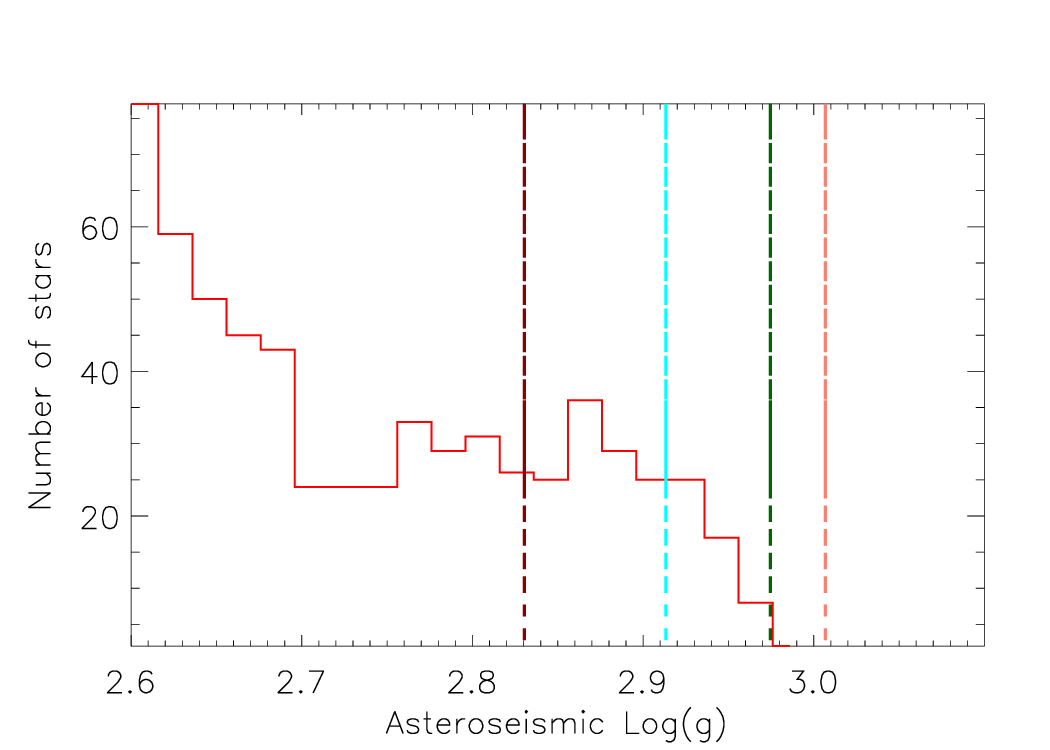}
  \caption{\textit{Top}: Seismic surface gravity (g) for the stars identified by seismology as belonging to the RC (in red) as a function of the spectroscopic surface gravity. The blue points correspond to the stars identified as RGB by spectroscopy on the contrary to seismic results. \textit{Bottom}: Seismic surface gravity for the same sample of seismic RC stars, focused on the bottom of the RC. The dashed lines correspond to the maximum values for secondary clump YREC models, with a core overshoot of $0.2$ pressure scale height. Dark red, cyan, dark green and salmon colors correspond, respectively, to stellar models with masses of $3.0$, $2.8$, $2.6$ and $2.4\Msol$.   
  \label{fig:Limit_clump}}
\end{figure}

 The spectroscopic and seismic comparison shows an overall good agreement (top part of Figure \ref{fig:Limit_clump}). However, it can be clearly seen that spectroscopic results gives $\logg$ values up to $3.2$, while asteroseismic results stop just below $3$. The discrepancies concern a few stars for which their spectroscopic values were overestimated compared to the seismic ones. This is in part due to an incorrect spectroscopic evolutionary status, as can be seen by the blue stars on the top part of Figure \ref{fig:Limit_clump}, which in turn caused the wrong surface gravity correction to be applied. The discrepancies between the two sets of values is apparent in the bottom part of Figure \ref{fig:Limit_clump} showing a distinct RC edge at $\logg$ $=$ $2.99\pm0.01$ for seismic results, on the contrary to spectroscopic measurements. This observed strong RC edge is consistent with previous models of secondary clump stars \citep{2011ApJS..192....3P,2018ApJ...868..150T} and can serve as an observational constraint for future stellar evolution models.
\newline

\begin{figure}                 % Insertion d'une figure = objet flottant
  \includegraphics[width=9cm]{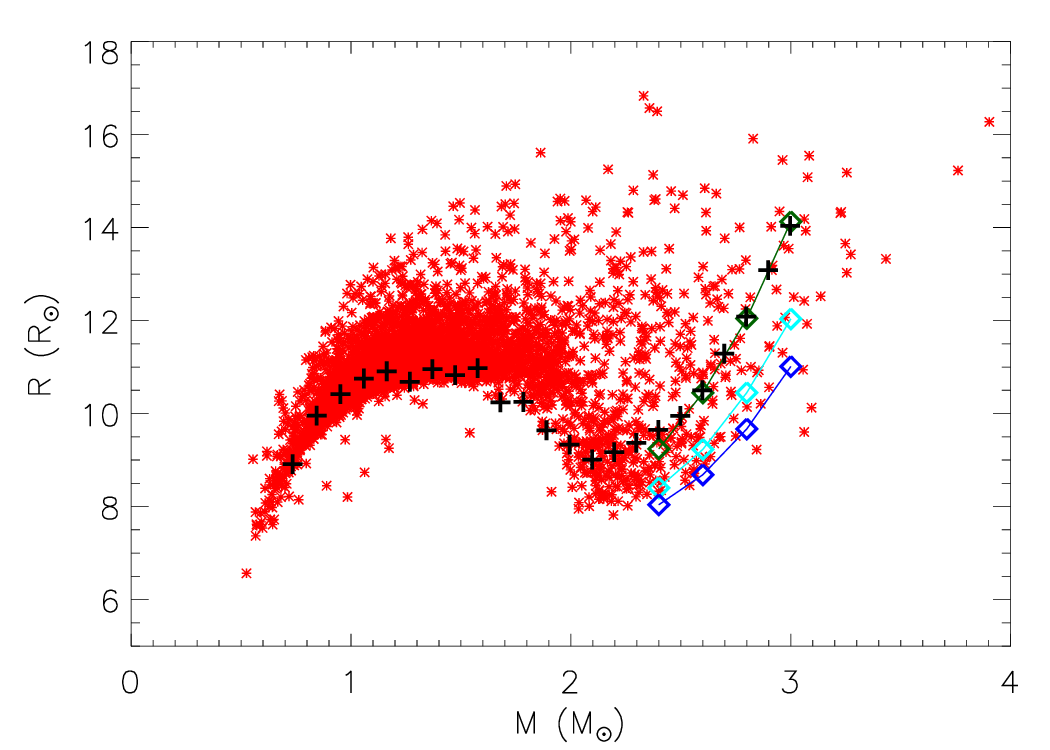}
  \caption{Seismic radius as a function of seismic mass for stars identified as RC following Section \ref{seismic_consensus}. The maximum $\logg$ values of the RC for the different YREC models are represented by the diamonds. The colors depict models with different core overshooting values: no overshooting for dark green, $0.1$ pressure scale height for cyan and $0.2$ pressure scale height for dark blue. The minimum radius values for the different MIST tracks are shown as black crosses. \label{fig:Radius_Mass}}
\end{figure}

In order to assess the possibilities to constrain models for such a clear limit of the secondary RC, we compare those results with models that were computed with the Yale Rotating Evolution Code \citep[YREC, ][]{1989ApJ...338..424P,2012ApJ...746...16V}. The details of the models are present in Section $2.2$ of \citet{2018ApJ...868..150T}. Several evolutionary tracks were computed with $4$ different masses ($2.4\Msol$, $2.6\Msol$, $2.8\Msol$, $3.0\Msol$) and $3$ different values of core overshooting (overshoot parameters of $0.0$, $0.1$ and $0.2$ pressure scale heights). Lower masses were not considered because of the difficulty of going beyond the Helium flash. The $\logg$ RC higher limit for the different models with highest overshooting value is shown on the bottom part of Figure \ref{fig:Limit_clump} and their minimum RC radius for each mass is depicted on Figure \ref{fig:Radius_Mass}.

From those Figures, it is clear that models without core overshooting or with low overshooting can not reproduce the observed results on the helium core mass. This conclusion was also reached previously to reproduce $\deltapi$ values by models for RC stars \citep{2015MNRAS.453.2290B,2015MNRAS.452..123C,2017MNRAS.469.4718B}. These previous works, showed that high core overshoot was necessary during the RC phase to match the observations. Here, it is only the model with the most amount of core overshooting until the beginning of the RC phase that can reach the base of the secondary clump in terms of $\logg$. Having such a hard limit for these RC stars can therefore bring important constraints on stellar models.
\newline

Finally, we can see on Figure \ref{fig:Radius_Mass} the sharp limit of the zero-age sequence of Helium burning stars (ZAHB) as pointed out by \citet{2021MNRAS.501.3162L} and \citet{2022NatAs...6..673L}. The presence of this sharp feature is also a confirmation of the robustness of the evolutionary status determination; the few number of stars with lower radius values compared to their masses is indeed consistent with previous results \citep{2022NatAs...6..673L}. To compare the position in radius of the ZAHB to theoretical prediction, we used the MESA Isochrones and Stellar Tracks \citep[MIST: ][]{2016ApJS..222....8D,2016ApJ...823..102C} based on the MESA models \citep{2011ApJS..192....3P,2013ApJS..208....4P,2015ApJS..220...15P,2018ApJS..234...34P}. We obtained tracks with masses between $0.8\Msol$ and $3\Msol$, solar metallicity and no extinction. In the MIST models, diffusive overshoot is included in their core and their envelope with an efficiency of $f\ind{ov,core}$ $=0.0160$ and $f\ind{ov,env}$ $=0.0174$ \citep{2016ApJ...823..102C,2000A&A...360..952H} The lower radius value during the He-burning phase are represented by black crosses in Figure \ref{fig:Radius_Mass} and reproduce well the global shape of the ZAHB. This curve shape is due to several phenomenon, the main one being that low-mass stars (M$\lsim$ $1.8\Msol$) start their Helium-burning phase with a degenerate core, therefore with the same He core mass \citep{1999MNRAS.308..818G,2013ApJ...766..118M}. The star with lower masses will then produce the same observed luminosity \citep{2016MNRAS.457L..59L}, therefore reach lower radius. For more massive stars, the He-burning starts in a non-degenerate core leading to lower helium core masses and radii \citep{1999MNRAS.308..818G,2013ApJ...766..118M}, thus producing the observed curved shape. Although the MIST tracks reproduce the global shape of the ZAHB, they do not reproduce the shape in detail. This could be because the full sample has a range of metallicities, and we have shown only the solar metallicity results and one type of overshooting. In order to test the influence on metallicity on the ZAHB shape, we restricted the mass and radius determination to stars with similar metallicities. It appears that it does not have an effect on the lower limit of the ZAHB. Therefore, only the difference in physical processes in the models can explain the differences between the observed ZAHB lower limit and the MIST models. We can also see that the MIST tracks are in agreement with the YREC models with no overshooting, which is surprising because overshooting has been implemented in MIST tracks \citep{2016ApJ...823..102C,2022ApJ...927...31T}. However, several other parameters have an influence on the models parameters for red giant stars, the main ones being the description of convective processes at the convective boundaries, the rate of specific nuclear reactions and mass loss at the tip of the RGB \citep[see, for a complete review,][]{2016ARA&A..54...95G}. A global comparison between theoretical models would be necessary to fully understand which parameters affect the differences between the models for massive stars, but is beyond the scope of this paper. Nonetheless, this analysis between theoretical models and observations shows that these results can be used to study the physics of RC models.

\section{Asymptotic giant branch stars classification}\label{AGB_classification}

The separation between RGB and AGB stars is difficult to perform with spectroscopy or seismology because the stellar properties of those red giants are very similar. However, two seismic techniques has claimed to be able to disentangle the two evolutionary status using the pressure mode pattern of the star oscillation spectra. Each of them expanded a technique that was tested and calibrated on the separation of RGB and RC stars. \citet{2012A&A...541A..51K} is based on the variation of the pressure mode frequencies as a function of frequency due to discontinuities present in the star's envelope while \citet{2019A&A...622A..76M} is based on the amplitude of the envelope autocorrelation function (EACF) used to measure the large separation $\Dnu$. The use of those two techniques simultaneously would therefore allow the separation between AGB and RGB stars with more certainty. However, \citet{2021A&A...650A.115D} observed that the agreement between those two techniques is present at high $\Dnu$ and $\numax$ but lowers quickly when the star evolves along the AGB or the upper part of the RGB. They found $35\%$ disagreement between the two techniques for stars with $\Dnu\leq1\mu$Hz. Because the beginning of the AGB phase starts with a $\Dnu\sim2.5\mu$Hz for a typical $1\Msol$ star, this discrepancy appears to rise rapidly. We, thus, need a new method to confirm that we can seismically disentangle the two evolutionary statuses.

With the current data set we possess, there are $76$ stars identified as AGB by $\Msix$ for which an evolutionary status is also given by $\Mdeux$. Between those two techniques, we only have a $48.7\%$ agreement, which correspond to a random distribution, therefore pointing toward a lack of agreement between the two methods. However, most of the stars for which the disagreement occurs have very low $\Dnu$. Although the stars with $\Dnu\leq1\mu$Hz show a systematic disagreement, the agreement fraction increases towards larger $\Dnu$, reaching $63.3\%$ for stars with $\Dnu >1.5\mu$Hz. However, this number is less than two standard deviations away from a random distribution for the star's identification. The small number of stars in that sample can explain why we have a discrepancy with previous results. It follows that, even if these results seems to be in poor agreement with the previous work of \citet{2021A&A...650A.115D}, we can not reach definitive conclusions about the efficiency of the methods from them. These disagreements confirms previous results that highlighted the need of an independent determination of the evolutionary status for high-luminosity giants, as stated in the previous paragraph.
\newline

By contrast, spectroscopy can be a useful discriminant between AGB and RGB in some cases. In much the same way as the stars on the RC are hotter than those on the RGB, we expect stars on the AGB to be hotter than those on the RGB at the same mass, metallicity, and surface gravity. AGB stars should also only be present above the luminosity of the RC. In order to have an independent analysis, we perform a spectroscopic determination of the evolutionary status of low surface gravity ($\logg$ $< 2.2$) using APOGEE data and evolutionary tracks. For each star in the APOKASC-3 sample, an interpolation with the model grid described by \citet{2017ApJ...840...17T} is performed using the seismic mass, metallicity, [M/H], alpha elements abundance ([$\alpha$/M]) and seismic $\logg$. The $\Teff$ that the interpolated model possess for each star, named afterwards $\Tint$, is then taken as a reference effective temperature. The measured offsets between the observed temperature $\Teff$ and the model-predicted temperature $\Tint$ were plotted as a function of metallicity, and similar to the results of \citet{2017ApJ...840...17T}, a trend between the temperature offsets and the stellar metallicity was detected. For this work, such a trend is not of scientific interest and so we define a function to remove this trend given the stellar metallicity, with the offset being $140$K if [Fe/H] $\geq$ $0.0$ and $190$[Fe/H] + $140$K if [Fe/H] $<$ $0.0$. We note that there have been suggestions that the trend should also be a function of alpha abundance \citep{2018A&A...612A..68S}, mass \citep{2018ApJ...856...10J}, surface gravity \citep{2012ApJ...755L..12B} and so forth (see \citet{2023Galax..11...75J} for a more complete discussion). However, we do not explore such complexities here, opting instead for a more simple fit to propose the technique, and expecting later authors to improve upon our results. 

$\DeltaTeff$ is thus obtained with the following relation:

\begin{equation}
\label{Eq:echelle_law}
    \DeltaTeff = \Teff - \Tint -190\mathrm{[Fe/H]} -140,
\end{equation}

\noindent
for stars with [Fe/H] $<$ $0.0$, and 

\begin{equation}
\label{Eq:echelle_law}
    \DeltaTeff = \Teff - \Tint -140,
\end{equation}

\begin{figure}                 % Insertion d'une figure = objet flottant
  \includegraphics[width=9cm]{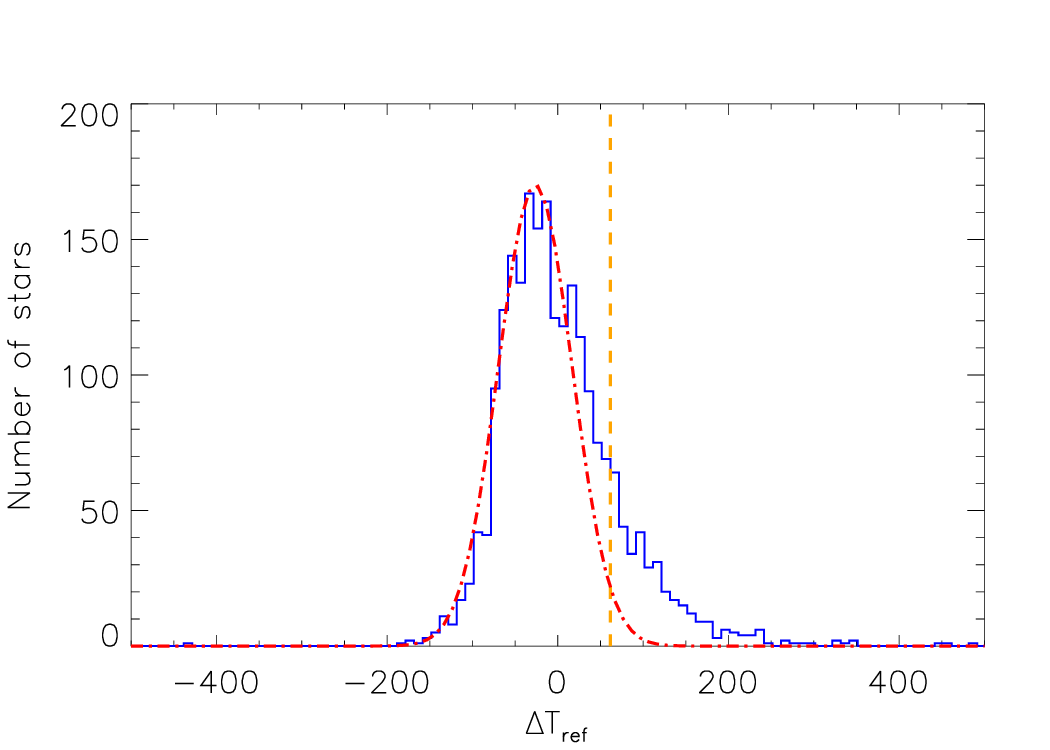}
  \caption{Distribution of stars identified as RGB or AGB, with $\logg$ $< 2.2$ (stars above the main RC population in a Kiel diagram), as a function of $\DeltaTeff$ (in blue). The dot-dashed red line shows a Gaussian fit of width $44.04$K centered on $-26.40$ following the $\alpha$-rich stars distribution. The orange dashed line indicate the two-$\sigma$ limit of the measured RGB sample distribution following Section \ref{Section:Spectro_Determination_technique}. \label{fig:Distribution_AGB}}
\end{figure}

\noindent
for stars with [Fe/H] $>$ $0.0$. $\DeltaTeff$ should therefore correspond to how much the effective temperature of the star diverges from model predictions, corrected for a metallicity dependent $\Teff$ offset in the models. The stars that show a deviation towards high effective temperature, accounting for the uncertainties on $\Teff$, should correspond to the AGB candidates. The $\DeltaTeff$ distribution can be seen on Figure \ref{fig:Distribution_AGB} where a clear asymmetry is present towards hotter stars. This feature in the distribution is a signature of the presence of AGB targets. We decided to take the separation between RGB stars and AGB candidates at $2$-$\sigma$ from the original RGB stars distribution as described in Section \ref{Section:Spectro_Determination_technique}. We also note that the $\Tref$ distribution is not exactly centered on $0$. In order to take this bias into account we selected $\alpha$-rich stars in our sample of low $\logg$ targets. For these stars, there is a well-defined median mass at a given metallicity, which allows us to better separate the AGB and RGB stars in the HR diagram \citep{2024arXiv241000102P}. We fitted the $\alpha$-rich RGB targets $\Tref$ distribution with a Gaussian and found a zero reference point of $-26.40\pm1.25$K. We then used this value as the mean of the original RGB gaussian distribution and select the AGB candidates at $2$-$\sigma$ from the original RGB distribution. The separation between RGB and AGB candidates is shown by the orange dashed line on Figure \ref{fig:Distribution_AGB} at $\Tref = 61.68$K. $398$ stars are identified as likely belonging to the AGB following this criterion. We would expect $96$ of them to be RGB stars scattered into the AGB domain by observational errors.

We note that, for the spectroscopic AGB candidates, the offset from the RGB locus tends to be slightly clearer before the metallicity-dependent offset is removed, suggesting that these stars might have metallicity-dependent model offsets that are different from the RGB stars. We also note that there are no obvious chemical differences between our spectroscopic AGB candidates and the RGB stars nearby (e.g. in carbon or oxygen), although we do not necessarily expect any such trends below the AGB bump \citep{2005MNRAS.356L...1S,2022ApJ...939...50C}, and we have not done an exhaustive search for more subtle signals. Some of these stars have low mass, which could indicate mass loss in previous phases of evolution.

%We also see the overdensity that we associate with the AGB bump in the population of high-alpha stars, which represents a smaller range of initial masses and a less complicated star formation history, suggesting that our detection is not the result of complex population and measurement effects.

\begin{figure}                 % Insertion d'une figure = objet flottant
  \includegraphics[width=9cm]{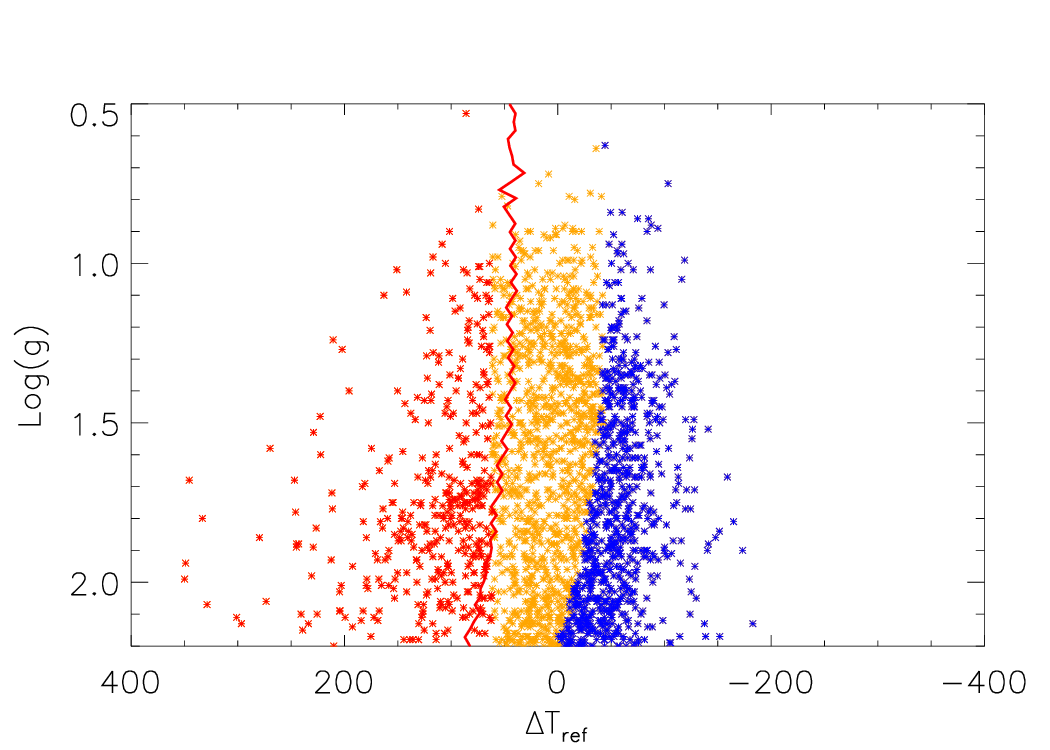}
  \includegraphics[width=9cm]{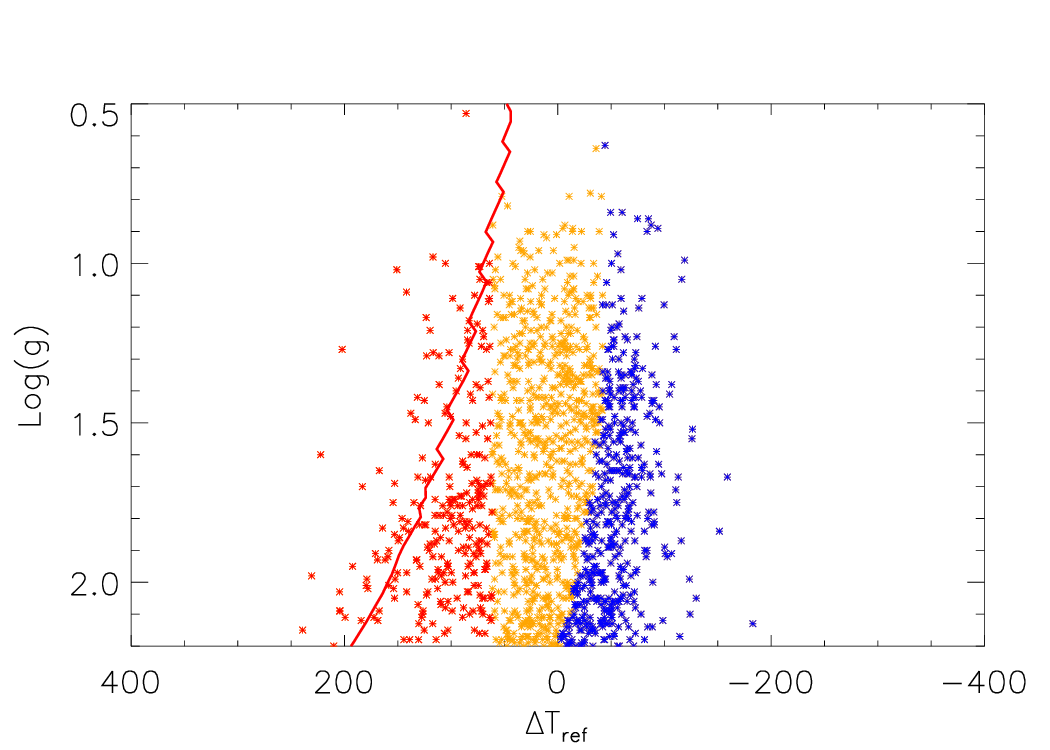}
  \includegraphics[width=9cm]{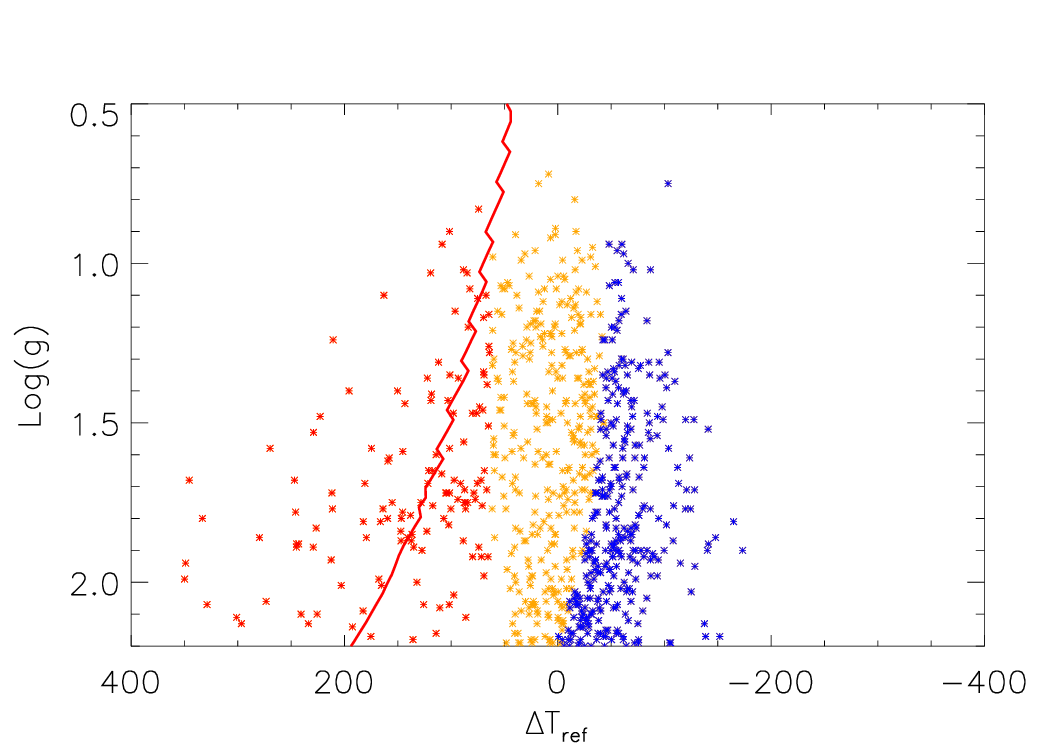}
  \caption{$\logg$ as a function of $\DeltaTeff$ for the stars above the main RC population. The blue points correspond to the stars identified as RGB targets, the orange points represent stars where the evolutionary status is uncertain and the red points are AGB candidates. \textit{Top:} full APOKASC-3 sample. The overplotted solid red line shows AGB-RGB $\Teff$ offset for the $1.7\Msol$ MIST evolutionary track. \textit{Middle:} $\alpha$-poor targets. The overplotted solid red line shows AGB-RGB $\Teff$ offset for the $1.0\Msol$ MIST evolutionary track. \textit{Bottom:} $\alpha$-rich targets. The overplotted solid red line shows AGB-RGB $\Teff$ offset for the $1.0\Msol$ MIST evolutionary track. \label{fig:true_RGB_selection}}
\end{figure}

\begin{figure}                 % Insertion d'une figure = objet flottant
  \includegraphics[width=9cm]{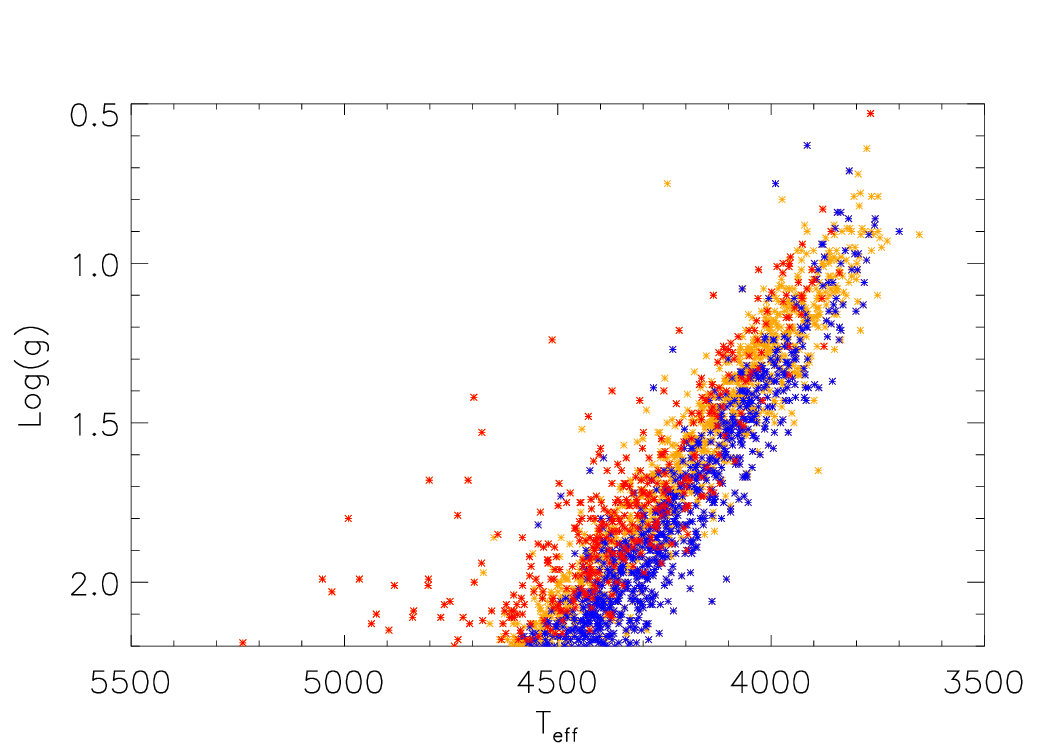}
  \includegraphics[width=9cm]{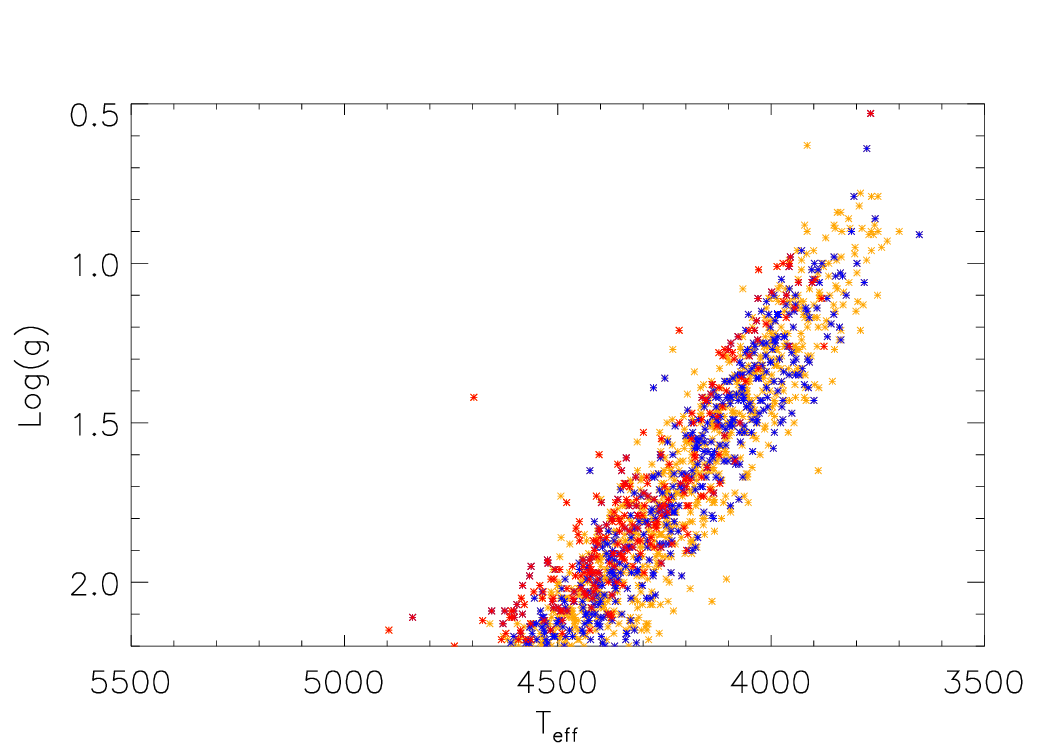}
  \includegraphics[width=9cm]{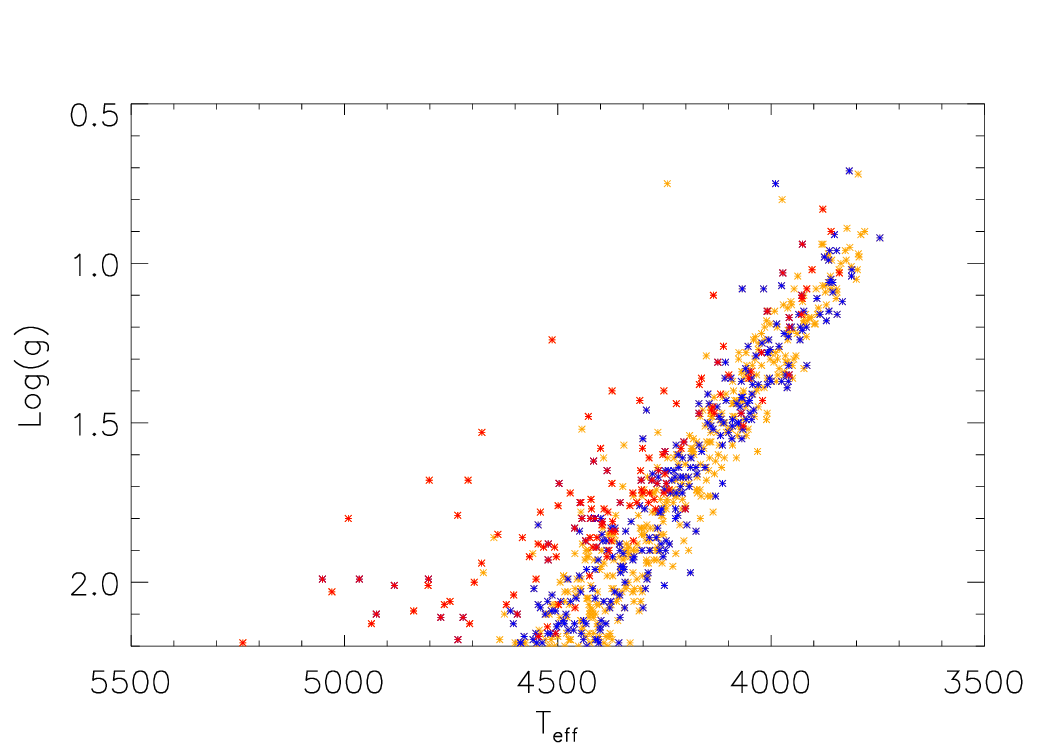}
  \caption{$\logg$ as a function of $\Teff$ for the stars above the main RC population. The blue points correspond to the stars identified as RGB targets, the orange points represent stars where the evolutionary status is uncertain and the red points are AGB candidates. \textit{Top:} full APOKASC-3 sample. \textit{Middle:} $\alpha$-poor targets. \textit{Bottom:} $\alpha$-rich targets.  \label{fig:true_selection_HR_diagram}}
\end{figure}

In this work, we also wanted to select stars for which the RGB evolutionary status was certain. In order to achieve that, we used the MIST models previously mentioned in Section \ref{Clump_limit} and computed the $\Teff$ difference between the AGB and RGB tracks at the same $\logg$. This $\Teff$ offset from RGB models rapidly decreases when we reach low $\logg$ as can be seen in the top panel of Figure \ref{fig:true_RGB_selection} for the $1.0\Msol$ MIST evolutionary track and in the other panels of that Figure for the $1.7\Msol$ MIST evolutionary track. The rapidity of the decrease highly depends on the model's mass, being stronger for higher mass stars. In order to be sure to select practically exclusively RGB stars, we took as a reference the $1.7$ MIST evolutionary track because its $\Teff$ offset value is lower than for lower-mass evolutionary tracks and only few targets have a higher mass in this part of the HR diagram for our stellar sample. We selected stars more than $2$-$\sigma$ away from that track following the original clump star distribution we measured in Section \ref{Section:Spectro_Determination_technique}. The stars not selected as AGB candidates or RGB are considered as having an undetermined status: RGB/AGB. The classification of the stars (RGB, candidate AGB and AGB/RGB) can be retrieved in \citet{2024arXiv241000102P}. It is also present in an adjoining file for which the first lines can be seen in Table \ref{Table_example_AGB_RGB_results}.

In Figure \ref{fig:true_RGB_selection} we show the $\DeltaTeff$ distribution for $\alpha$-poor (middle panel) and $\alpha$-rich stars (bottom panel) exclusively. The $\alpha$-rich stars regrouping targets with similar ages and metallicity, we then expect them to show a clear difference of $\DeltaTeff$ between the AGB and RGB targets. In the bottom part of Figure \ref{fig:true_RGB_selection} we can indeed see a separation between a group of stars near $\DeltaTeff = 0$, corresponding to RGB targets, and a hotter group of stars. This separation is mainly visible for stars with $\logg$ $> 2.0$. The hotter group of stars corresponds to the MIST $1.0\Msol$ AGB evolutionary track and can therefore easily be identified with AGB targets. We can also point out that the AGB selection we performed corresponds to the observed separation between the two groups of stars, therefore confirming that our spectroscopic AGB determination is consistent. In the middle panel of Figure \ref{fig:true_RGB_selection} ($\alpha$-poor stars) we can also note the absence of the hottest stars present in the other panels. They correspond to old low-mass $\alpha$-rich targets and have likely suffered heavy mass loss during the RGB phase. We also plotted the identified AGB and RGB targets in the HR diagram (Figure \ref{fig:true_selection_HR_diagram}). In this configuration, the separation between AGB and RGB targets is less clear. However we can see that the colder the star, the more likely it is identified as belonging to the RGB, and the hotter the star, the most likely it is identified as belonging to the AGB. It shows also that the hottest stars correspond to the $\alpha$-rich targets and are all identified as belonging to the AGB.
\newline

\begin{table*}[h]
\centering
\begin{tabular}{ccccccc}

%add in the discussion the number of quarters so we know which light-curves are high-quality.

KIC number &  $\Teff$ (K) & $\sigma_{\Teff}$ (K) & $\logg$ & $\sigma_{\ind{\logg}}$ & $\Tref$ (K) & AGB/RGB evolutionary status \\
\hline
  $10001167$   &  $4558$   &  $84$  &  $2.18$  &  $0.07$  &  $-4.25$ &  $1$ \\
  $10004975$   &  $4099$   &  $69$  &  $1.35$  &  $0.05$  &  $101.12$ &  $2$ \\
  $10023978$   &  $4230$   &  $74$  &  $1.60$  &  $0.06$  &   $61.53$ &  $0$ \\
  $10025788$   &  $3853$   &  $65$  &  $0.88$  &  $0.05$  &   $-6.66$ &  $0$ \\
\hline
\end{tabular}
\caption{First lines of a joined file summarizing the AGB or RGB evolutionary status determination. The different columns correspond respectively to the KIC number, $\Teff$, $\sigma_{\Teff}$, $\logg$, $\sigma_{\ind{\logg}}$, $\Tref$ and the AGB/RGB classification ($0$ for AGB/RGB, $1$ for certain RGB and $2$ for candidate AGB). \label{Table_example_AGB_RGB_results}}
\centering
\end{table*}

\begin{figure}                 % Insertion d'une figure = objet flottant
  \includegraphics[width=9cm]{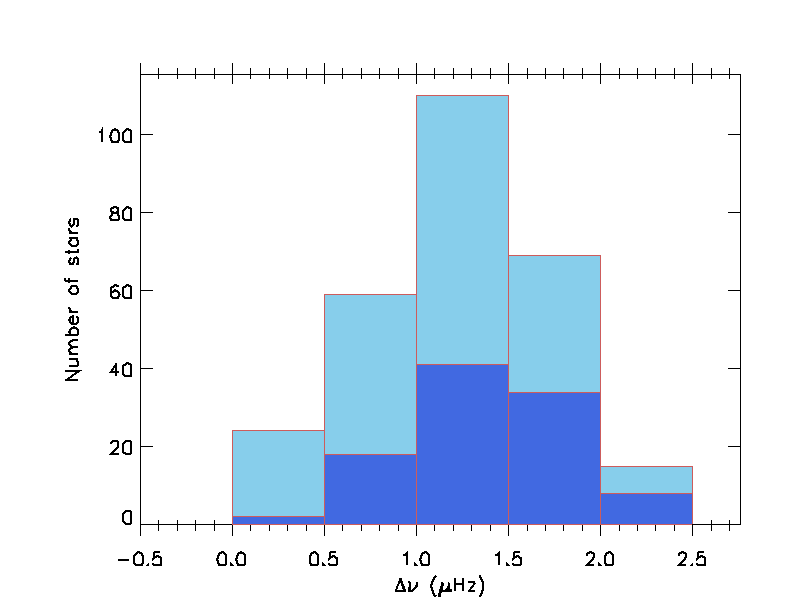}
  \includegraphics[width=9cm]{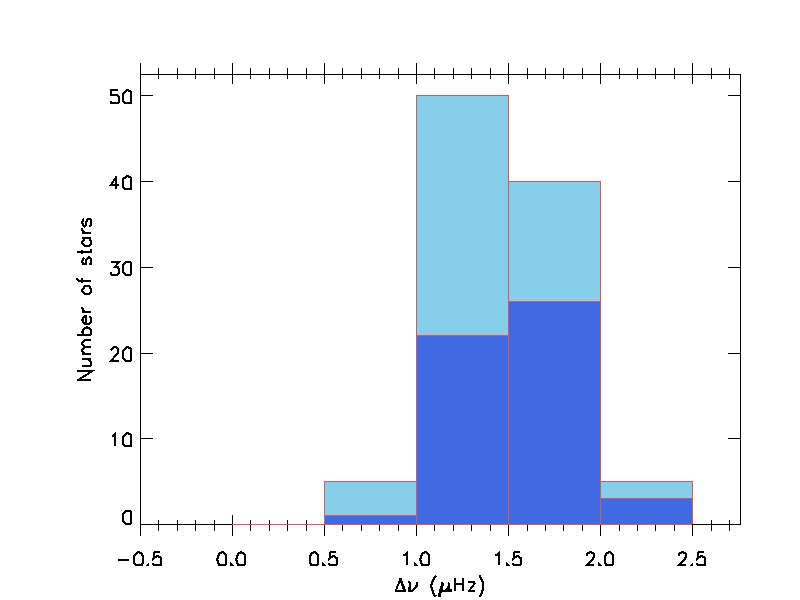}
  \caption{Distribution of stars spectroscopically identified as belonging to the AGB and for which an AGB evolutionary status is given by $\Mdeux$ (top) and by $\Msix$ (bottom) as a function of $\Dnu$. The dark blue histogram represent the star for which the classification agrees with the seismic technique while the whole sample is shown in cyan. \label{fig:histo_AGB_comparison}}
\end{figure}

Within this spectroscopic AGB candidate star sample, we identified $104$ stars for which an evolutionary status is determined with the seismic method of \citet{2012A&A...541A..51K}. Among those, $52.0\%$ are in agreement with the spectroscopic identification of AGB targets. Concerning the method of \citet{2019A&A...622A..76M}, $266$ stars correspond to the AGB candidate spectroscopic sample. $37.2\%$ are in agreement with the AGB candidates identified by our spectroscopic determination. The agreement between the spectroscopic and seismic results are particularly low and can correspond to a random identification, even if we consider that the number of false positive determination of AGB targets is non-negligible and that the number of AGB is expected to be significantly lower compared to RGB targets \citep{2022MNRAS.509.4430K}. In order to investigate this behavior, we plot in Figure \ref{fig:histo_AGB_comparison} the agreement between the spectroscopic and the seismic methods as a function of $\Dnu$. We can see that the consensus between the two types of techniques is rapidly degrading at low $\Dnu$, below $70\%$ for stars with $\Dnu <2\mu$Hz. Therefore, we can not assess that one or the other seismic or spectroscopic method is performing better for the evolutionary status determination at the exception of the vicinity of the RC where the three methods gives similar results. With these results, we can say that disentangling AGB and RGB with asteroseismology is limited to the stars close to the RC characteristics (with $\Dnu \sim 2\mu$Hz). Again, the small sample of targets doesn't allow us to reach firm conclusions and can explain why our results differ from previous ones. Nonetheless, we would advise to be cautious when extrapolating the efficiency of one of those technique for stars with a large separation below $2\mu$Hz.
\newline

\section{Conclusion}

The determination of red giant evolutionary status is of primary importance for stellar population, stellar interiors and galactic archaeology studies. In this paper, we created the most precise and up-to-date red giant evolutionary status catalog for $\Kepler$ asteroseismic targets to be used in the future on stellar evolution works and stellar population analysis. This catalog was obtained with asteroseismic and spectroscopic data. Several automated asteroseismic evolutionary status determination methods were combined to obtain the final sample separating RC and RGB stars. The different techniques were found in very good agreement with each other, with the ones using the mixed-mode pattern the most accurate but delivering fewer results in general. This determination was compared to the evolutionary status estimation given by spectroscopic results from the APOKASC-3 $\Kepler$ sample, which correspond to a subsample of the red giant $\Kepler$ legacy sample. We, then, focus on this subsample for a more detailed analysis. The spectroscopic determination was shown to be in very good agreement with the asteroseismic results, therefore proving the robustness of the calibrated spectroscopic evolutionary status determination. The stars for which a disagreement arises between the spectroscopic and asteroseismic classification appeared to have peculiar characteristics for their evolutionary status (high-mass on the RGB, high $\logg$ for clump targets or stars at the beginning of the AGB branch), therefore explaining why a confusion can arise between the two evolutionary status. This new catalog is in agreement with previous work and can be used to further test stellar models, including the necessary assumptions on the overshoot amount. We demonstrated that the RC has a clear observable edge situated at $2.99\pm0.01$ in $\logg$. In agreement with previous results, we identified and confirmed the minimum radius of the zero-age sequence of Helium burning stars (ZAHB), leaving only a few stars with a radius lower than this observed limit. Those stars, as pointed out by \citet{2022NatAs...6..673L}, likely correspond to peculiar red giants who experienced mass loss during their evolution. Finally, we tested the possibility to disentangle RGB from AGB stars on the upper part of the giant branch with asteroseismology and spectroscopy. We found that the seismic methods are in agreement with the spectroscopic data on the identification of AGB targets but only in the vicinity of the RC, for stars with $\Dnu \sim 2\mu$Hz. The results between the different techniques diverge rapidly when the star evolves along the AGB showing that identifying these stars remains difficult. It shows that the seismically identified AGB targets needs to be used with caution regarding the certainty of their evolutionary status for future works.

\begin{acknowledgements}
 M.V acknowledges support from NASA grant 80NSSC18K1582 and funding from the European Research Council (ERC) under the European Union’s Horizon 2020 research and innovation programme (Grant agreement No. 101019653). R.~A.~G. acknowledges the support of the PLATO and GOLF grants from the Centre National d’Études Spatiales (CNES). This Paper includes data collected by the \emph{Kepler} mission and obtained from the MAST data archive at the Space Telescope Science Institute (STScI). Funding for the \emph{Kepler} mission is provided by the NASA Science Mission Directorate. STScI is operated by the Association of Universities for Research in Astronomy, Inc., under NASA contract NAS 5–26555. M.H. acknowledges support from NASA grant 80NSSC24K0228. D.S. is supported by the Australian Research Council (DP190100666). S.M.\ acknowledges support by the Spanish Ministry of Science and Innovation with the Ramon y Cajal fellowship number RYC-2015-17697, the grant number PID2019-107187GB-I00, the grant no. PID2019-107061GB-C66, and through AEI under the Severo Ochoa Centres of Excellence Programme 2020--2023 (CEX2019-000920-S).
\end{acknowledgements}

\bibliographystyle{aa} % style aa.bst
\bibliography{evolution}

\appendix

\listofobjects
\end{document}